\documentclass[aps,pre,showpacs,amsmath,amssymb,amsfonts,lengthcheck,twocolumn,longbibliography]{revtex4-2}
\usepackage{graphicx}
\usepackage{subfigure}
\usepackage{amsthm}
\usepackage{verbatim}
\usepackage{dcolumn}% Align table columns on decimal point
\usepackage{bm}% bold math
\usepackage{color}
\usepackage[colorlinks=true,citecolor=blue,linkcolor=blue,urlcolor=blue]{hyperref}%
\usepackage{xcolor}
\usepackage{dsfont}
\newcommand{\bra}[1]{\left\langle #1\right|}
\newcommand{\ket}[1]{\left|#1\right\rangle}
\newcommand{\braket}[2]{\left\langle #1|#2\right\rangle}

\newcommand{\tr}[1]{\mathrm{tr}\left\{#1\right\}}

\newcommand{\la}{\left\langle}
\newcommand{\ra}{\right\rangle}

\newcommand{\e}[1]{\exp{\left(#1\right)}}

\newcommand{\bla}{bla\\bla\\bla\\bla\\bla}

\newcommand{\mrm}[1]{\mathrm{#1}}

\begin{document}

\title{Bosons Outperform Fermions -- The Thermodynamic Advantage of Symmetry}
\date{\today}
\author{Nathan M. Myers}
\email{myersn1@umbc.edu}
\author{Sebastian Deffner}
\email{deffner@umbc.edu}
\affiliation {Department of Physics, University of Maryland Baltimore County, Baltimore, MD 21250, USA}

\begin{abstract}
We examine a quantum Otto engine with a harmonic working medium consisting of two particles to explore the use of wave function symmetry as an accessible resource. It is shown that the bosonic system displays enhanced performance when compared to two independent single particle engines, while the fermionic system displays reduced performance. To this end, we explore the trade-off between efficiency and power output and the parameter regimes under which the system functions as engine, refrigerator, or heater. Remarkably, the bosonic system operates under a wider parameter space both when operating as an engine and as a refrigerator.
\end{abstract}

\maketitle

\section{Introduction} 

With the widespread adoption of the steam engine during the industrial revolution, thermodynamics emerged as a physical theory that could describe and optimize the performance of these devices \cite{kondepudi}. While modern thermodynamics has expanded far beyond its original scope, heat engines have remained the canonical systems for studying thermodynamic mechanisms. Not only do they have clear practical applications, but they also provide a paradigmatic way of studying how the thermodynamic properties of a system evolve -- with applications ranging from biological processes, over climate systems, to black holes \cite{muller, shaw, johnson}. 

Quantum systems, subject to inherent fluctuations and decidedly non-equilibrium in nature, introduce new challenges for applying the framework of thermodynamics \cite{deffner}. Nevertheless, quantum heat engines \cite{deffner,scovil} provide a natural foundation for studying thermodynamic behavior in quantum systems in comprehensible terms. For instance, heat can always be found as the change in energy during an isochoric stroke, just as work can be found from the change in energy during an isentropic stroke \cite{callen}.    

This might explain the plethora of studies to investigate possible enhancements of engine performance through the exploitation of quantum resources including coherence \cite{stwo, sthree, Uztwo, wattwo, dann, Feldtwo, Watanabe, hardal}, measurement effects \cite{buffoni}, squeezed reservoirs \cite{Niedenzu, long, rosstwo}, quantum phase transitions \cite{ma}, and quantum many-body effects \cite{Chen, hardal, Beau, Li}. Other works have examined the fundamental differences between quantum and classical thermal machines \cite{Quantwo, gardas, Friedenberger}, finite time cycles \cite{Cavina, Feldtwo, zhengthree}, utilizing shortcuts to adiabaticity \cite{abahfour, abahthree, abahfive, Beau, Campo, Funotwo, dann, Li}, operating over non-thermal states \cite{Yunger, cherubim}, non-Markovian effects \cite{Uzdin}, magnetic systems \cite{Munoztwo, Munoz, Penathree, Penatwo, Pena, Penafive}, anharmonic potentials \cite{Zhengtwo}, optomechanical implementation \cite{Zhang}, quantum dot implementation \cite{Penatwo, Munoz, Penafive}, implementation in 2D materials \cite{Pena, Munoz}, classical engines coupled to quantum systems \cite{Scully}, quantum cooling \cite{Levytwo, ntwo}, relativistic systems \cite{Penafour, Munozthree}, degeneracy effects \cite{Barrios, Penathree}, and autonomous cycles \cite{roulet}. Moreover, recent experimental advances have demonstrated the practical implementation of nanoscale heat engines \cite{rossnagel, josefsson}, including those that harness quantum resources \cite{Klaers, Horne}.

Two primary quantities of practical interest when characterizing the performance of a heat engine are its efficiency and power output. However, analysis of an ideal engine assumes that the strokes of the engine are carried out quasistatically over an infinitely slow period, maximizing efficiency but resulting in zero power output \cite{callen}. Rather, one is interested in the efficiency at maximum power (EMP) \cite{Curzon}. To this end, it has been shown that such analyses are particularly fruitful for quantum engines \cite{Kosloff, Geva, Feldmann, Esposito, Rezek, etwo, dtwo, Uzthree, abah, rosstwo}.      

In this paper we explore a quantum Otto cycle similar to the model pioneered by Kosloff \cite{Kosloff}. Our working medium consists of two particles in a harmonic trap. We examine performance, including efficiency, power, EMP, trade-off between efficiency and power, and the parameter regimes where the cycle functions as various types of thermal machines, depending on if the particles are bosons, fermions, or non-interacting and independent (often referred to as ``classical"). Through this, we explore the effect of wave function symmetry on engine performance. We find that in all examined performance characterizations the bosons perform better in comparison to the independent particles, while the fermions perform worse. Symmetry effects in engine performance have been explored for other potentials or interactions \cite{huang, Zheng, jaramillo}, however, our work provides additional insight through a fully analytical model of the complete dynamics, demonstrating how these effects arise solely from the underlying wave function symmetry.           

\section{Preliminaries}

\subfigcapmargin = 10cm
\begin{figure*}
\centering
\subfigure[]{
\includegraphics[width=.3\textwidth]{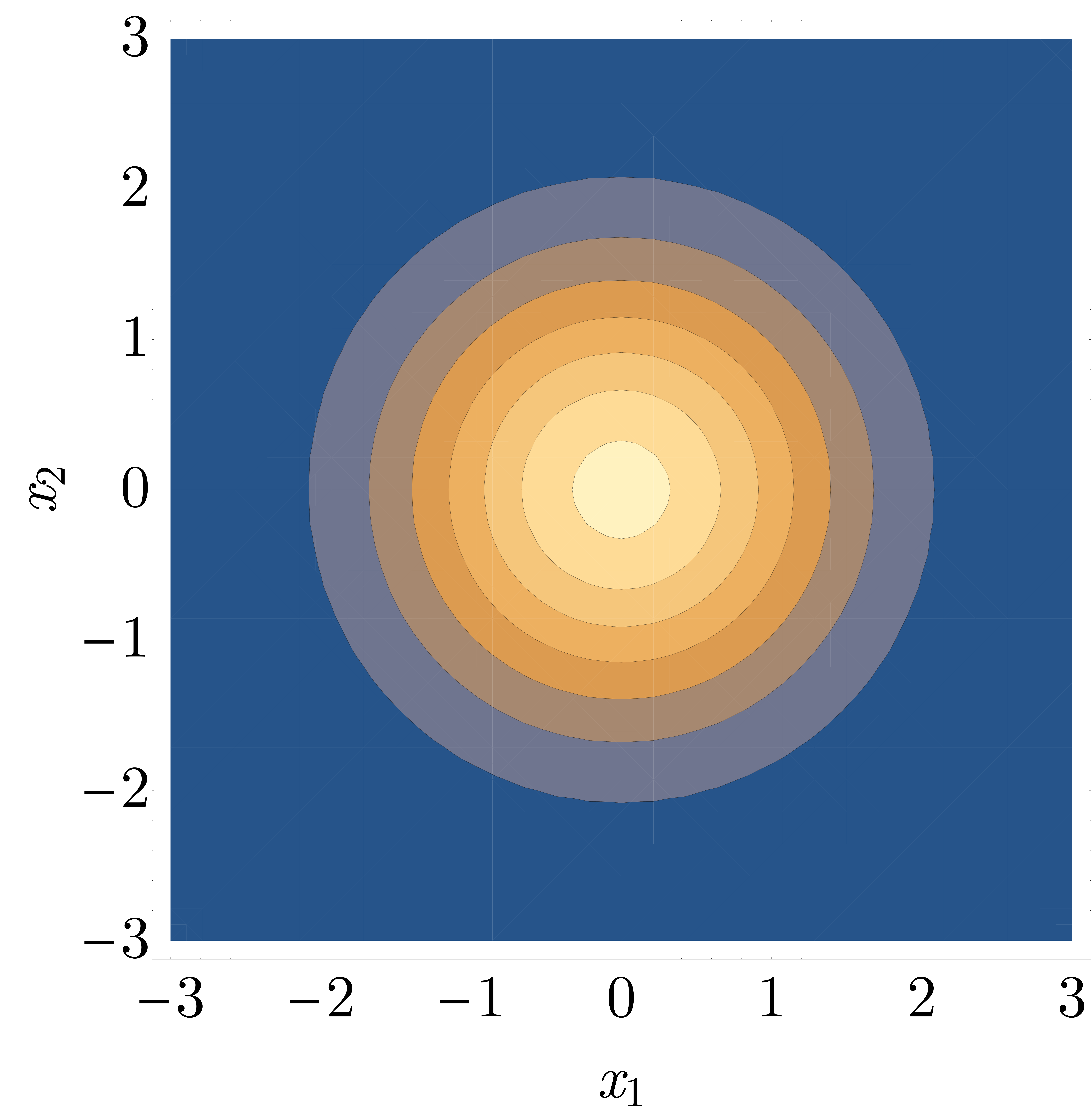}
}
\subfigure[]{
\includegraphics[width=.3\textwidth]{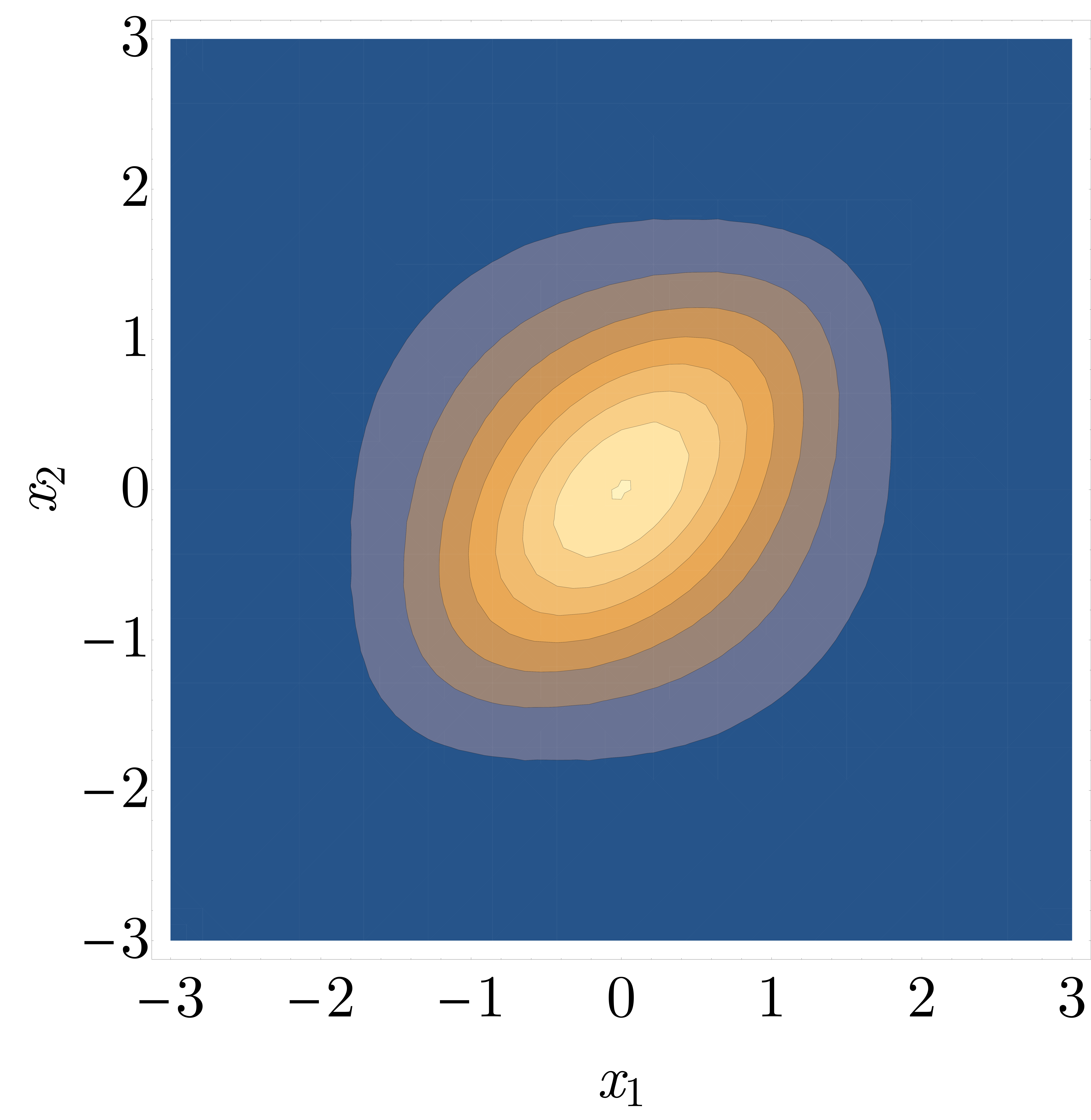}
}
\subfigure[]{
\includegraphics[width=.3\textwidth]{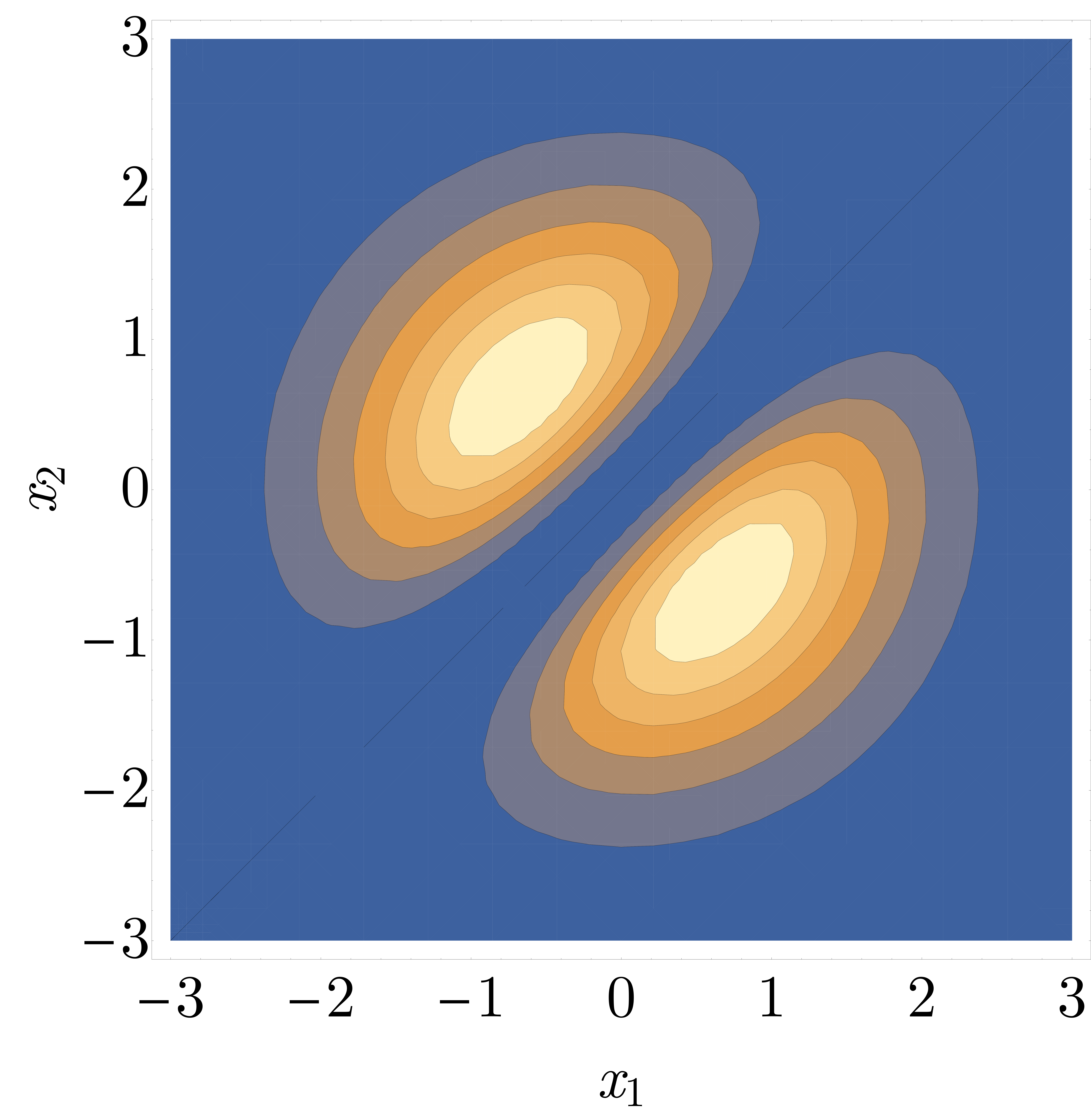}
}
\caption{\label{fig:thermal} Position distributions of the thermal state \eqref{eq:thermal} of (a) two independent particles, (b) two bosons, and (c) two fermions in a harmonic potential. Parameters are $\hbar = k_B = 1$, and $\beta = \omega = m = 1$.}
\end{figure*}

Our working medium is given by two noninteracting, spinless particles, either both bosons or both fermions. Excluding any additional interaction terms and considering spinless particles (or alternatively particles of identical spin) allows us to isolate behavior arising from symmetry effects. The potential is a harmonic trap, such as a linear Paul trap, whose frequency can be varied with time. This is a two-particle generalization of the experimental model proposed in Ref.~\cite{abah}.   

To analyze an engine operating in finite-time, we need the dynamics of a quantum harmonic oscillator with a time-dependent angular frequency $\omega (t)$ that varies from $\omega_1$ at $t=0$ to $\omega_2$ at $t=\tau$. To this end, we start by briefly reviewing the case of a single particle, which we can then generalize to the case of two particles. 

For a single particle the Hamiltonian is the sum of the kinetic and potential energies,
\begin{equation}
H = \frac{p^2}{2m}+\frac{1}{2} m \omega^2 (t) x^2.
\end{equation}
Following the method developed by Husimi \cite{husimi} we can solve the time-dependent Schr\"odinger equation by introducing the Gaussian wave-function ansatz,
\begin{equation}
\psi_t (x) = \e{\frac{i}{2 \hbar} (a_t x^2 + 2 b_t x + c_t)},
\end{equation}
where $a_t$, $b_t$, and $c_t$ are time-dependent coefficients. This allows us to reduce the Schr\"odinger equation to a set of three coupled differential equations for $a_t$, $b_t$, and $c_t$ that can be solved by mapping to the equation of motion of the classical time-dependent harmonic oscillator,
\begin{equation}
\label{eq:classical}
\ddot{X}_t + \omega^2(t) X_t = 0.
\end{equation}

The single-particle propagator then reads \cite{husimi},
\begin{equation}
U_1= \sqrt{\frac{m}{2 \pi i h X_t}}\e{\frac{im}{2\hbar X_t} (\dot{X_t}x^2-2xx_0+Y_tx_0^2)}\,,
\end{equation}  
where $X_t$ and $Y_t$ are time-dependent solutions to Eq.~\eqref{eq:classical} with initial conditions $X_0 = 0$, $\dot{X}_0 = 1$ and $Y_0 = 1$, $\dot{Y}_0 = 0$.

This framework can be directly expanded to two particles with Hamiltonian $H_\mrm{tot} = H_1 + H_2$. For two particles the pure state wave function,
\begin{equation}
\Psi_{n_1,n_2}(x_1,x_2) = \frac{1}{2} \left[ \psi_{n_1}(x_1) \psi_{n_2}(x_2) \pm \psi_{n_1}(x_2) \psi_{n_2}(x_1) \right]\,,  
\end{equation} 
consists of the symmetric (for bosons) or antisymmetric (for fermions) linear combination of the single particle wave functions. 

\begin{widetext}
Accordingly, the two-particle equilibrium thermal state reads,
\begin{equation}
\label{eq:thermal}
\rho_0(x_1,x_2, y_1,y_2) = \frac{1}{Z} \sum_{n_1 = 0}^{\infty} \sum_{n_2 = 0}^{\infty} \e{-\beta \hbar \omega (n_1 + n_2 + 1)}\Psi_{n_1,n_2}^*(x_1,x_2)  \Psi_{n_1,n_2}(y_1,y_2).
\end{equation}
where $Z$ is the standard partition function $Z = \tr{ \e{-\beta H}}$. Inserting the harmonic oscillator energy eigenstates in position representation for $\psi_{n_1}$ and $\psi_{n_2}$ yields the position space density operator, the full expression of which is given in Appendix \ref{Appendix A}.
\end{widetext}

The thermal state \eqref{eq:thermal} already displays notable differences in behavior arising from the wave function symmetry. This can be most easily observed from the states' Wigner quasi-probability distributions \cite{wigner}.  See Appendix \ref{Appendix C} for the full expressions. By integrating the Wigner distributions over the momentum components we determine the position probability distributions for each thermal state. Figure~\ref{fig:thermal} depicts the position distribution for two independent particles, two bosons, and two fermions. The stretching of the boson distribution along the diagonal (where the position of the two particles coincide) is a demonstration of the effective attraction (boson bunching), conversely, the splitting of the fermion distribution along its diagonal is a demonstration the corresponding effective repulsion between fermions  (Pauli exclusion principle).

A heat engine cycle will necessarily involve some compression and expansion of the working medium. Considering this, the differences in the thermal state from the exchange forces already give us reason to suspect that symmetry should affect engine performance.  

\begin{widetext}
Finally, to determine the time evolved density operator we further require the proper two particle evolution operator. It can be shown that in energy representation the two particle propagator is the symmetric (for bosons) or antisymmetric (for fermions) linear combination of single particle propagators \cite{gong}. The same is true in position representation (see Appendix \ref{Appendix B} for a full derivation),
\begin{equation}
U_2(x_1,x_1^0,x_2,x_2^0) = \frac{1}{2} \left[  U_1(x_1,x_1^0)\, U_1(x_2,x_2^0)\pm U_1(x_1,x_2^0)\, U_1(x_2,x_1^0)\right]\,,
\end{equation} 
where again the plus is for bosons and the minus for fermions.
\end{widetext}

\section{Finite-Time Quantum Otto Cycle}

\begin{figure}
\includegraphics[width=.48\textwidth]{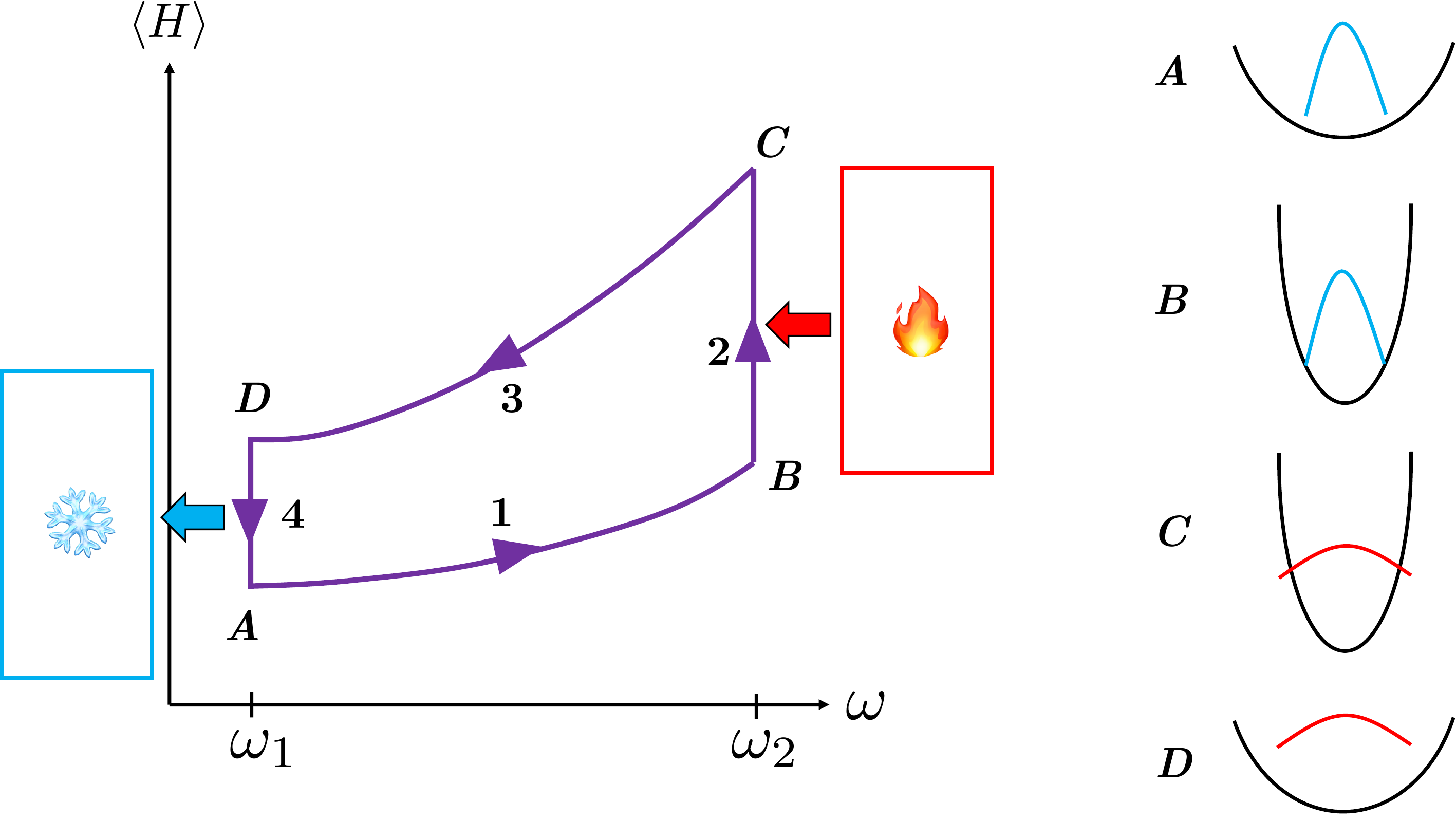}
\caption{\label{fig:cycle} Energy-frequency diagram of a quantum Otto cycle with representation of each stroke's trapping potential.}
\end{figure}
Classically, the Otto cycle consists of four strokes (1) isentropic compression, (2) isochoric heating, (3) isentropic expansion, and (4) isochoric cooling \cite{callen}. Note that the classical and the quantum Otto cycle are implemented in a fundamentally different manner: Typically, in the quantum Otto cycle, the isentropic strokes are given by unitary strokes \cite{Kosloff,deffner}, such that the von Neumann entropy remains constant. This is in contrast to classical adiabatic strokes which are carried out quickly  to prevent heat transference, and hence keep the thermodynamic entropy constant \cite{callen}.

In our model the oscillator frequency plays the role of volume, with the compression and expansion strokes corresponding to closing (increasing frequency) and opening the trap (decreasing frequency), respectively. The heating (cooling) stroke corresponds to coupling the oscillator respectively to a high (low) temperature bath that increases (decreases) the energy of the system \cite{abah}. This thermodynamic cycle is illustrated graphically in Fig.~\ref{fig:cycle}: At $A$ the system is in an equilibrium thermal state with inverse temperature $\beta_1$, and frequency $\omega_1$. The isentropic compression stroke is carried out via unitary evolution to non-thermal state $B$ with increased frequency $\omega_2$. The system is then coupled to the hot reservoir and allowed to thermalize to state $C$ at inverse temperature $\beta_2$ and frequency $\omega_2$. The frequency is then decreased unitarily during the expansion stroke resulting in state $D$ with $\omega_1$ and $\beta_2$. Finally, the system is coupled to the cold reservoir and allowed to thermalize back to $\beta_1$, returning it to its original state $A$.

Applying the two particle propagator derived above to the thermal state density operator we can determine the state of the system after either of the unitary strokes (1 and 3). The full expression can be found in Appendix \ref{Appendix B}.   Using the explicit expression for the time evolved state \eqref{eq:W_evolved}, we obtain the internal energy  at each corner of the cycle,
\begin{equation}
\begin{split}
\la H \ra_A &= \frac{\hbar \omega_1}{2}\,\left( 3 \mathrm{coth}(\beta_1 \hbar \omega_1) + \mathrm{csch}(\beta_1 \hbar \omega_1) \mp 1 \right), \\
\la H \ra_B &= \frac{\hbar \omega_2}{2}\, Q^*_{12}\, \left( 3 \mathrm{coth}(\beta_1 \hbar \omega_1) + \mathrm{csch}(\beta_1 \hbar \omega_1) \mp 1 \right), \\
\la H \ra_C &= \frac{\hbar \omega_2}{2}\,\left( 3 \mathrm{coth}(\beta_2 \hbar \omega_2) + \mathrm{csch}(\beta_2 \hbar \omega_2) \mp 1 \right), \\
\la H \ra_D &= \frac{\hbar \omega_1}{2}\, Q^*_{21}\, \left( 3 \mathrm{coth}(\beta_2 \hbar \omega_2) + \mathrm{csch}(\beta_2 \hbar \omega_2) \mp 1 \right).
\end{split}
\end{equation}
Here $Q^*_{12}$ and $Q^*_{21}$ are dimensionless parameters that measure the degree of adiabaticity of the isentropic strokes \cite{husimi}. They are given by,
\begin{equation}
Q^*= \frac{1}{2 \omega_0 \omega_\tau}\left[\omega^{2}_0\left(\omega^{2}_\tau X_{\tau}^2 + \dot{X}_{\tau}^2\right)+\left(\omega^{2}_2 Y_{\tau}^2+\dot{Y}_{\tau}^2\right)\right]\,
\end{equation} 
where $X_t$ and $Y_t$ are solutions of Eq.~\eqref{eq:classical} with $\omega(t=0)=\omega_1$ and $\omega(t=\tau)=\omega_2$ during compression, and the reverse during the expansion. Note that for a completely adiabatic stroke $Q^* = 1$ and in general $Q^* \geq 1$ \cite{husimi,abah}. 

Since the only energy exchange during the isentropic strokes is in the form of work, we can determine the average work directly from the differences in internal energy $\la W_1 \ra= \la H \ra_B - \la H\ra_A$ and $\la W_3 \ra = \la H \ra_D - \la H \ra_C$. Analogously, we have from the isochoric strokes that $\la Q_2 \ra = \la H \ra_C - \la H \ra_B$ and $\la Q_4 \ra = \la H \ra_A - \la H \ra_D$.

\begin{widetext}
\section{Engine Characterizations} 

\subsection{Efficiency and Power}

\begin{figure*}
\centering
\subfigure[]{
\includegraphics[width=.48\textwidth]{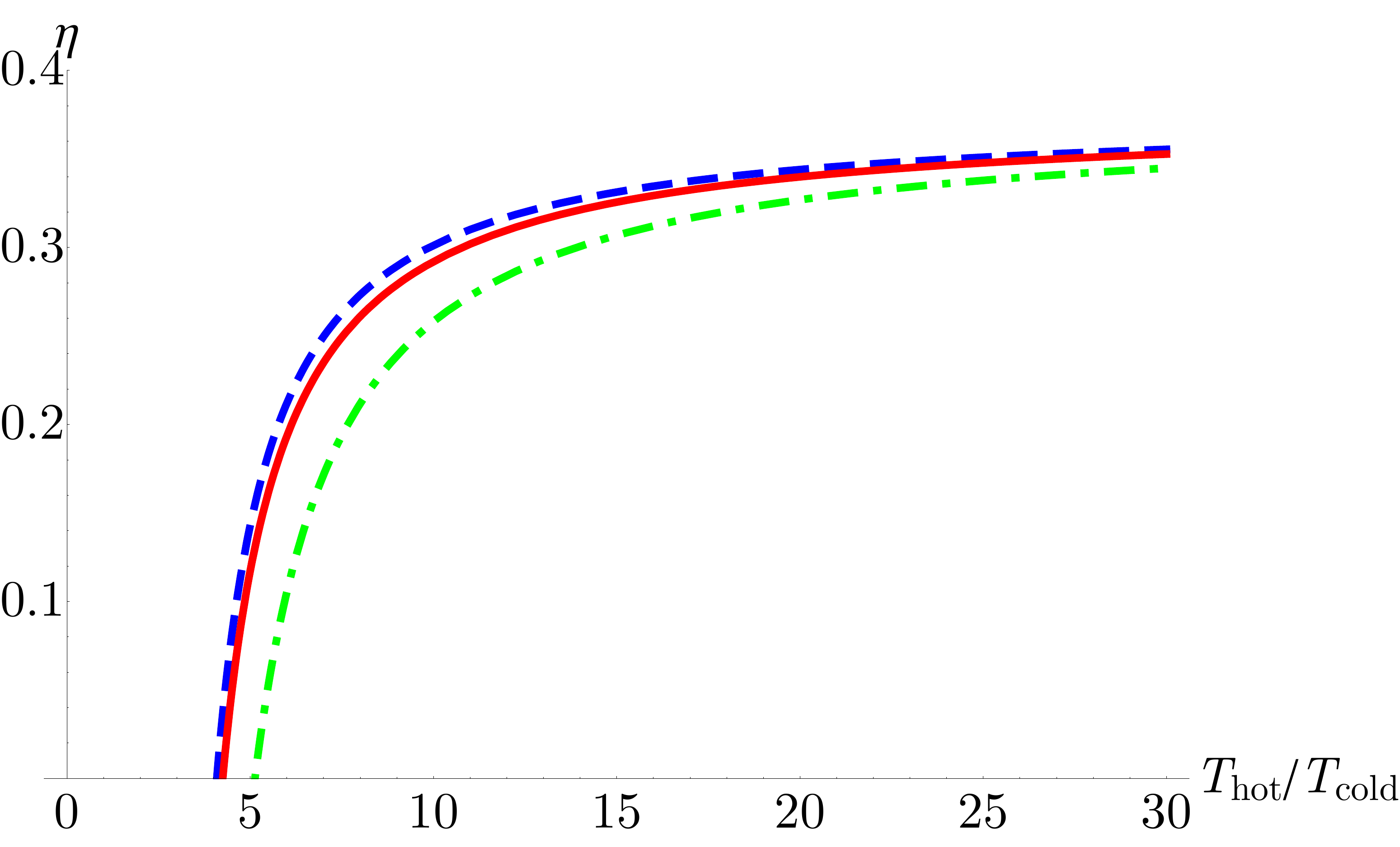}
}
\subfigure[]{
\includegraphics[width=.48\textwidth]{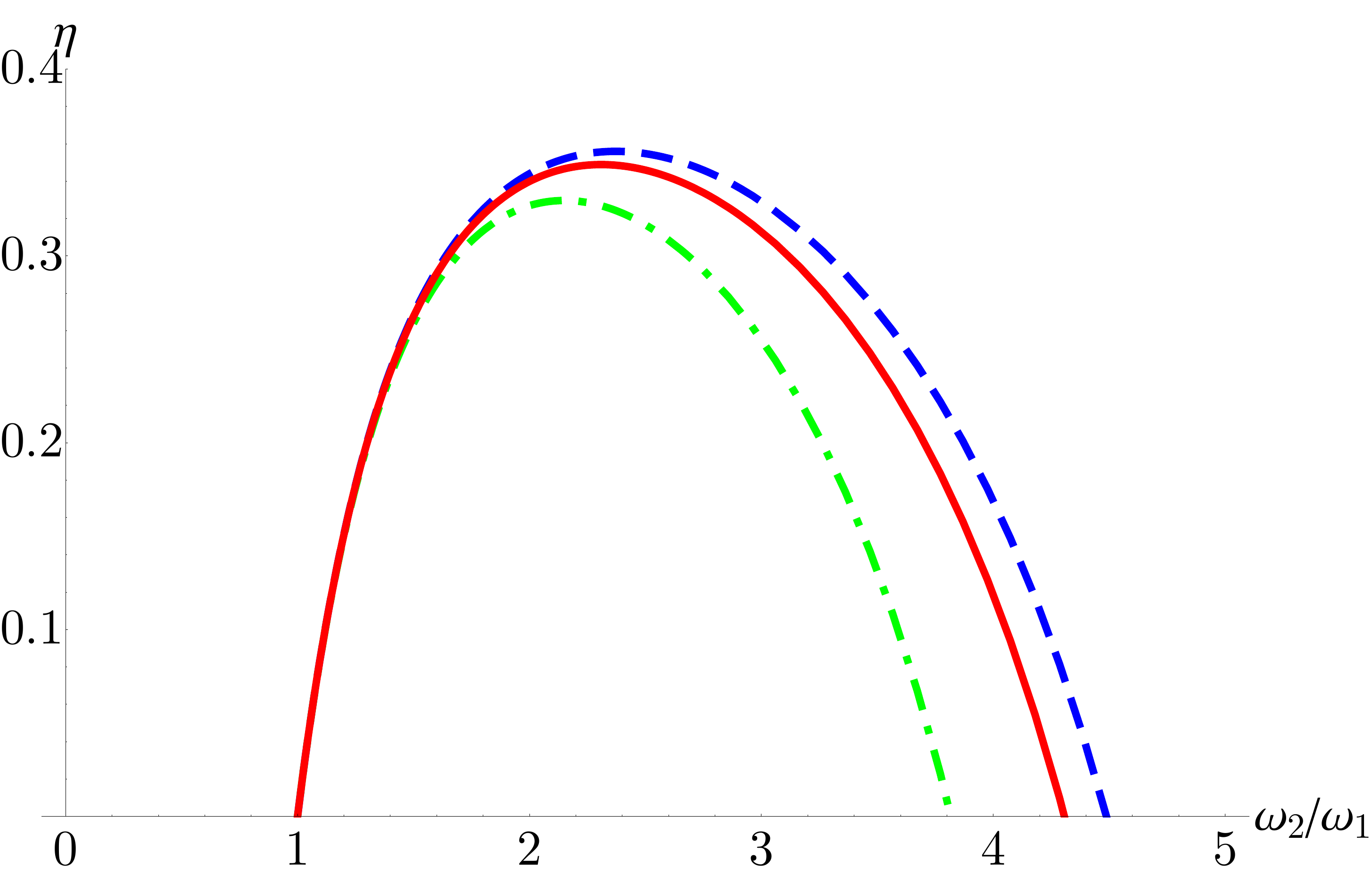}
}
\caption{\label{fig:eff}Efficiency as a function of the ratio of bath temperatures (a) and final to initial frequencies (b) for a bosonic engine (dashed, blue line), fermionic engine (dot-dashed, green line) and single particle engine (solid, red line). We have taken $\hbar = k_B = 1$. Other parameters are, for (a) $\omega_1 = 1$, $\omega_2=2$, and for (b) $T_{\mathrm{cold}}=1$, $T_{\mathrm{hot}}=20$.}
\end{figure*}

For any thermodynamic engine, the efficiency is defined as the ratio of the total work to the heat input while the power is defined by ratio of the total work to the cycle time,    
\begin{equation}
\eta = -\frac{\la W_1 \ra + \la W_3 \ra}{\la Q_2 \ra} \quad \text{and}\quad P = \frac{\la W_1 \ra + \la W_3 \ra}{\tau_1+\tau_2+\tau_3+\tau_4}\,.
\end{equation}
Thus, the full two-particle efficiency becomes,
\begin{equation}
\label{eq:eff}
\eta = 1 - Q^*_{21} \frac{\omega_1}{\omega_2} \mp \frac{\omega_1 \left(Q^*_{12}Q^*_{21}-1\right)\left(3 \mathrm{coth}(\beta_1 \hbar \omega_1) + \mathrm{csch}(\beta_1 \hbar \omega_1) \mp 1\right)}{\omega_2 \left(\pm 3 \mathrm{coth}(\beta_2 \hbar \omega_2) \mp Q^*_{12}(3 \mathrm{coth}(\beta_1 \hbar \omega_1) + \mathrm{csch}(\beta_1 \hbar \omega_1) \mp 1 ) \pm \mathrm{csch}(\beta_2 \hbar \omega_2)-1 \right)},
\end{equation}
where the top sign denotes the bosonic efficiency and the bottom sign the fermionic efficiency.
\end{widetext}

To check the consistency of our results, we examine the classical limit (high-temperature, quasi-static) of the maximum efficiency. In the quasi-static limit ($Q^*_{12}=Q^*_{21}=1$) Eq.~\eqref{eq:eff} reduces (as expected) to, 
\begin{equation}
\eta = 1 - \frac{\omega_1}{\omega_2}\,,
\end{equation} 
which is the quasistatic efficiency of the ideal quantum Otto engine \cite{Kosloff}.

\begin{figure*}
\centering
\subfigure[]{
\includegraphics[width=.48\textwidth]{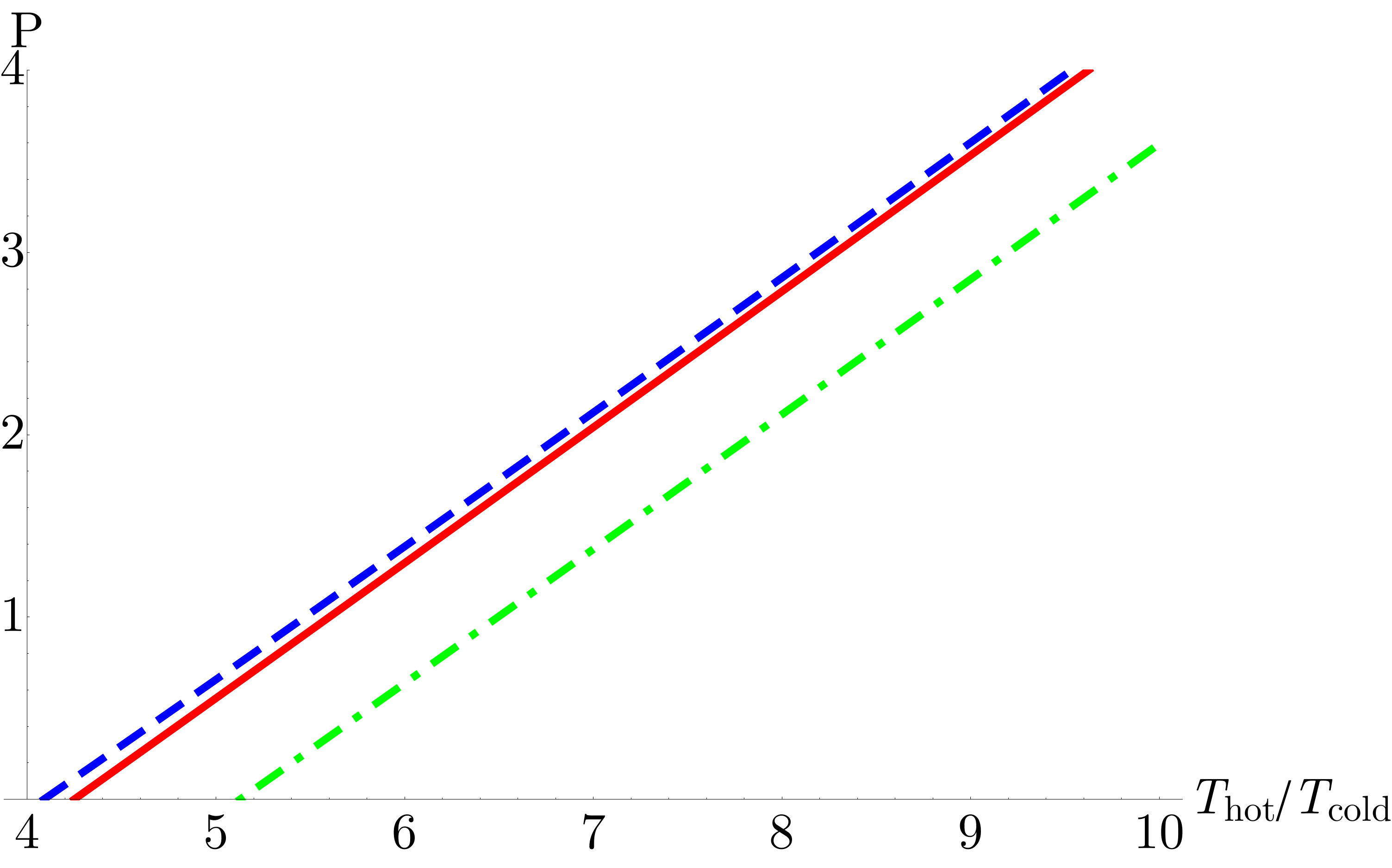}
}
\subfigure[]{
\includegraphics[width=.48\textwidth]{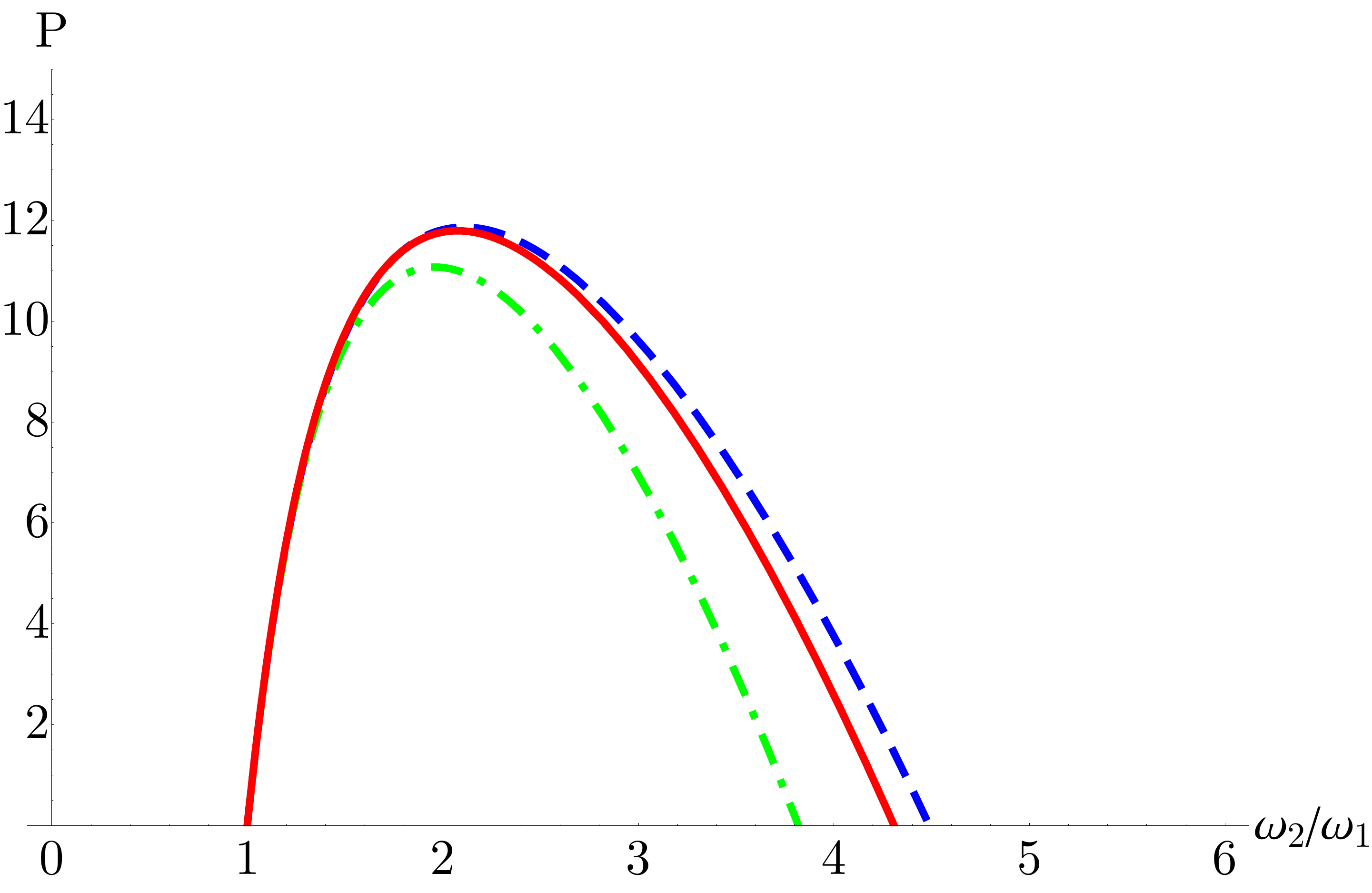}
}
\caption{\label{fig:power} Power as a function of the ratio of bath temperatures (a) and final to initial frequencies (b) for a bosonic engine (dashed, blue line), fermionic engine (dot-dashed, green line) and two independent single particle engines (solid, red line). We have taken $\hbar = k_B = 1$. Other parameters are, for (a) $\omega_1 = 1$, $\omega_2=2$, $\tau = 1$, and for (b) $T_{\mathrm{cold}}=1$, $T_{\mathrm{hot}}=20$, $\tau = 1$.}
\end{figure*}   

In order to restrict the operation of our cycle to the regime in which it behaves as an engine (and thus ensure discussion of efficiencies is valid) we impose the following (positive work) conditions,
\begin{equation}
\la Q_2 \ra > 0, \quad \la Q_4 \ra < 0, \quad \text{and}\quad \la W_{\mathrm{total}} \ra > 0. 
\end{equation}
In the classical limit these positive work conditions are equivalent to,
\begin{equation}
\frac{\beta_2}{\beta_1} \leq \frac{\omega_1}{\omega_2}. 
\end{equation}
Therefore, we see that, indeed, the maximum efficiency becomes,
\begin{equation}
\eta_\mrm{max} = 1 - \frac{T_c}{T_h}\,,
\end{equation}
which is just the Carnot efficiency.

To examine the behavior of the efficiency and power outside of limiting cases we first select a protocol $\omega (t)$. For the sake of simplicity, we start with the ``sudden switch" protocol, which corresponds to an instantaneous change from $\omega_1$ to $\omega_2$ \cite{abah}. In this case
\begin{equation}
Q^{*}_{12}=Q^{*}_{21}= \frac{\omega_2^2+\omega_1^2}{2 \omega_2 \omega_1}\,.
\end{equation}

The efficiency for bosonic and fermionic engines using this protocol is shown in comparison to that of a single particle quantum Otto engine in Fig.~\ref{fig:eff}. Note that efficiency is identical for the sum of any number of single particle engines. We see a notable enhancement in efficiency over the single particle engine at intermediate bath temperature ratios for the bosonic engine, and a universal decrease in efficiency for the fermionic engine. We see a similar enhancement for bosons and reduction for fermions at high frequency ratios.

The power output for bosonic and fermionic engines using this protocol is shown in comparison to that of the sum of two single particle quantum Otto engines in Fig.~\ref{fig:power}. Note that in the case of the sudden switch protocol the cycle time consists of just the sum of the thermalization times, as the isentropic strokes are considered to be instantaneous. Again, we see an enhancement to the power output in the case of bosons, and a significantly larger decrease in power output in the case of fermions, in comparison to the equivalent number of single particle engines.

\subsection{Efficiency at Maximum Power}

	\begin{figure}
		\includegraphics[width=.48\textwidth]{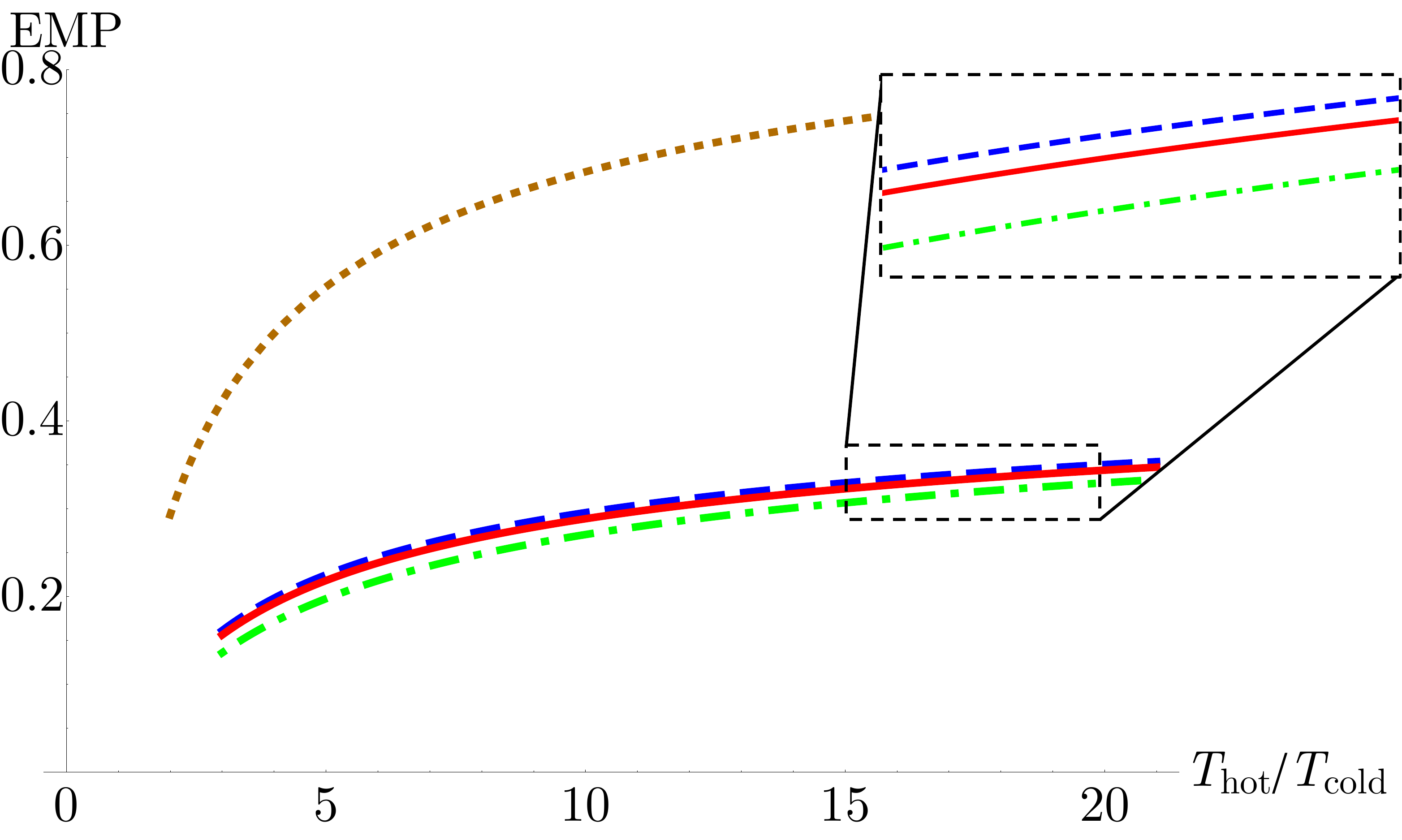}
		\caption{\label{fig:emp} EMP as a function of the ratio of bath temperatures for a bosonic engine (dashed, blue line), fermionic engine (dot-dashed, green line) and two independent single particle engines (solid, red line) given in comparison to the Curzon-Ahlborn efficiency (dotted, brown line). We have taken $\hbar = k_B = 1$. Other parameters are $\omega_1 = 1$ and $\tau=1$.}
	\end{figure}

As stated above, due to the inherent trade-off between efficiency and power, the more practically significant characterization of heat engine performance is the efficiency at maximum power. To determine this, we first maximize the power with respect to the second frequency, $\omega_2$, assuming $\omega_1$, the cycle time, and the bath temperatures are held fixed.

Carrying out this maximization in the classical limit yields,
\begin{equation}
\frac{\omega_2}{\omega_1} = \sqrt{\frac{\beta_1}{\beta_2}}.
\end{equation}
Thus, we have
\begin{equation}
\label{eq:CA}
\eta_{\mathrm{EMP}} = 1 - \sqrt{\frac{T_c}{T_h}}\,,
\end{equation}
which is nothing else but the Curzon-Ahlborn efficiency \cite{Curzon}. This is in full agreement with recent findings in Refs.~\cite{Rezek} and \cite{dtwo} that show in the quasistatic and classical high-temperature limits respectively, the EMP of a quantum harmonic Otto is given by Eq.~\eqref{eq:CA}

To examine the EMP outside of the classical limit we again choose the sudden switch protocol. Figure~\ref{fig:emp} shows the EMP as a function of the ratio of bath temperatures for our three working mediums. Here we see the same pattern as before, an enhancement over the equivalent number of single particle engines in the case of bosons, and a significantly larger reduction in the case of fermions. 
	
\subsection{Efficiency and Power Trade-off}

	\begin{figure}
		\includegraphics[width=.48\textwidth]{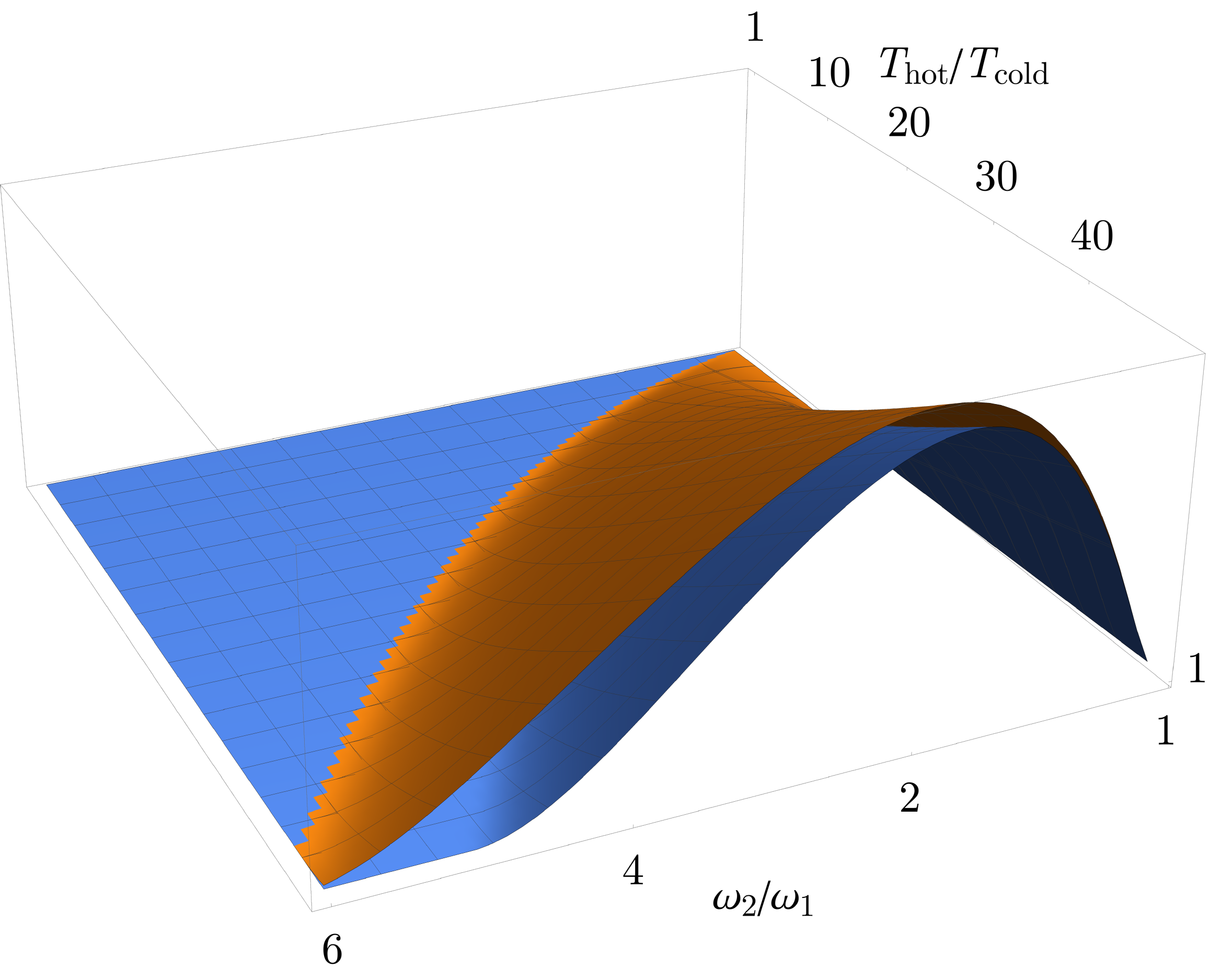}
		\caption{\label{fig:trade-off} Efficient power as a function of the ratio of both bath temperatures and initial and final frequencies for a bosonic engine (top, orange layer) and a fermionic engine (bottom, blue layer). We have taken $\hbar = k_B = 1$ and $\tau=1$.}
	\end{figure} 

While the EMP provides information about the amount of efficiency sacrificed to achieve \textit{maximum} power, it can be enlightening to also examine the trade-off across the entire parameter space. Various trade-off measures have been put forward (see Refs.~\cite{Angulo, hernandez, Pietzonka, Yilmaz, Arias, Velasco, Singh} for further discussion). Here we use the ``efficient power" criterion proposed by Yilmaz, which is defined as the simple product of efficiency and power \cite{Yilmaz}. This provides us with a direct measure of the power output gained per corresponding unit decrease in efficiency. The efficient power for both bosons and fermions for the case of the sudden switch protocol is shown in Fig.~\ref{fig:trade-off} as a function of both the ratio of bath temperatures and initial and final frequencies. We see that over the whole parameter space the efficient power for bosons is superior to the trade-off for fermions.  

\subsection{Operational Parameter Regimes}

	\begin{figure}
		\includegraphics[width=.48\textwidth]{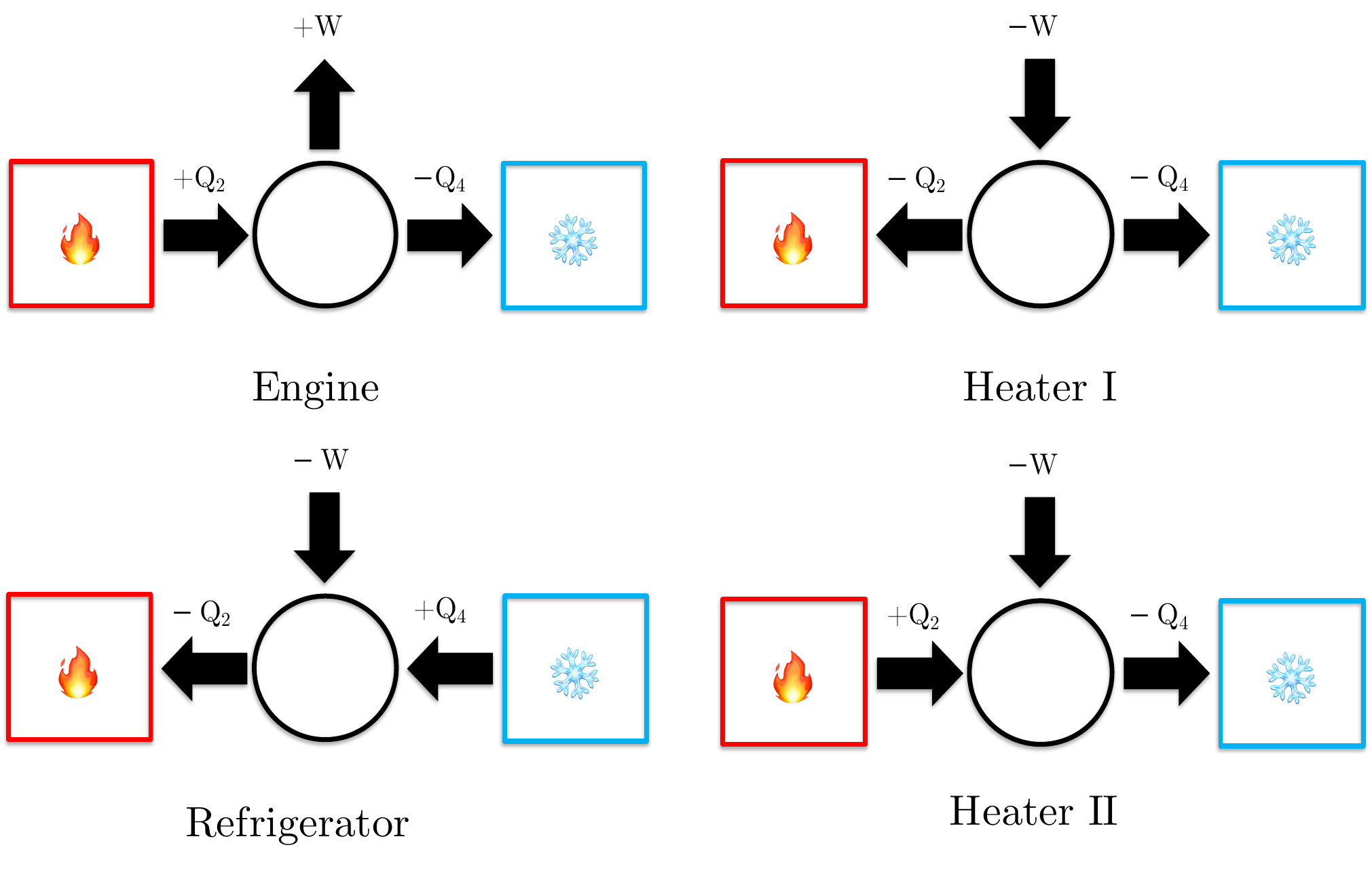}
		\caption{\label{fig:block}Block diagrams for all allowed thermal machines.}
	\end{figure}

\begin{figure*}
\centering
\subfigure[]{
\includegraphics[width=.3\textwidth]{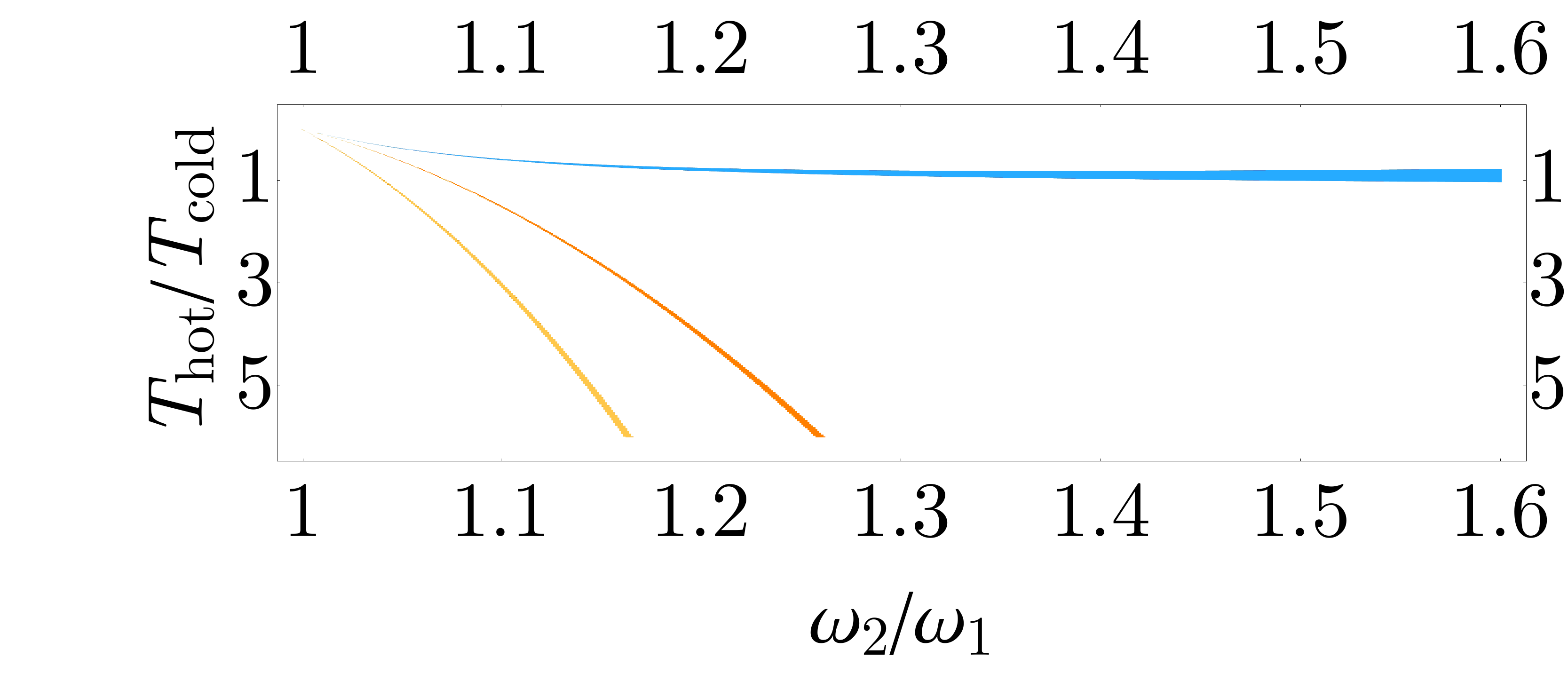}
}
\subfigure[]{
\includegraphics[width=.3\textwidth]{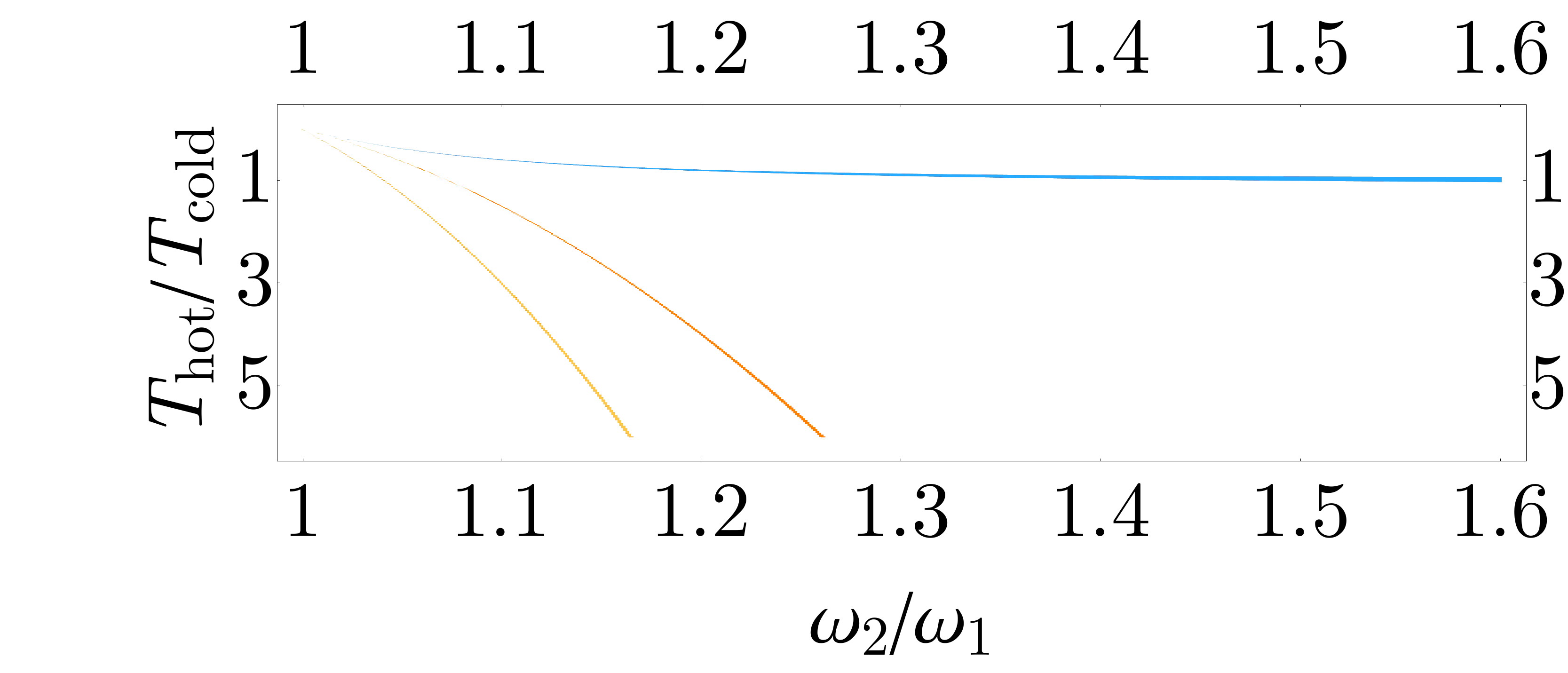}
}
\subfigure[]{
\includegraphics[width=.3\textwidth]{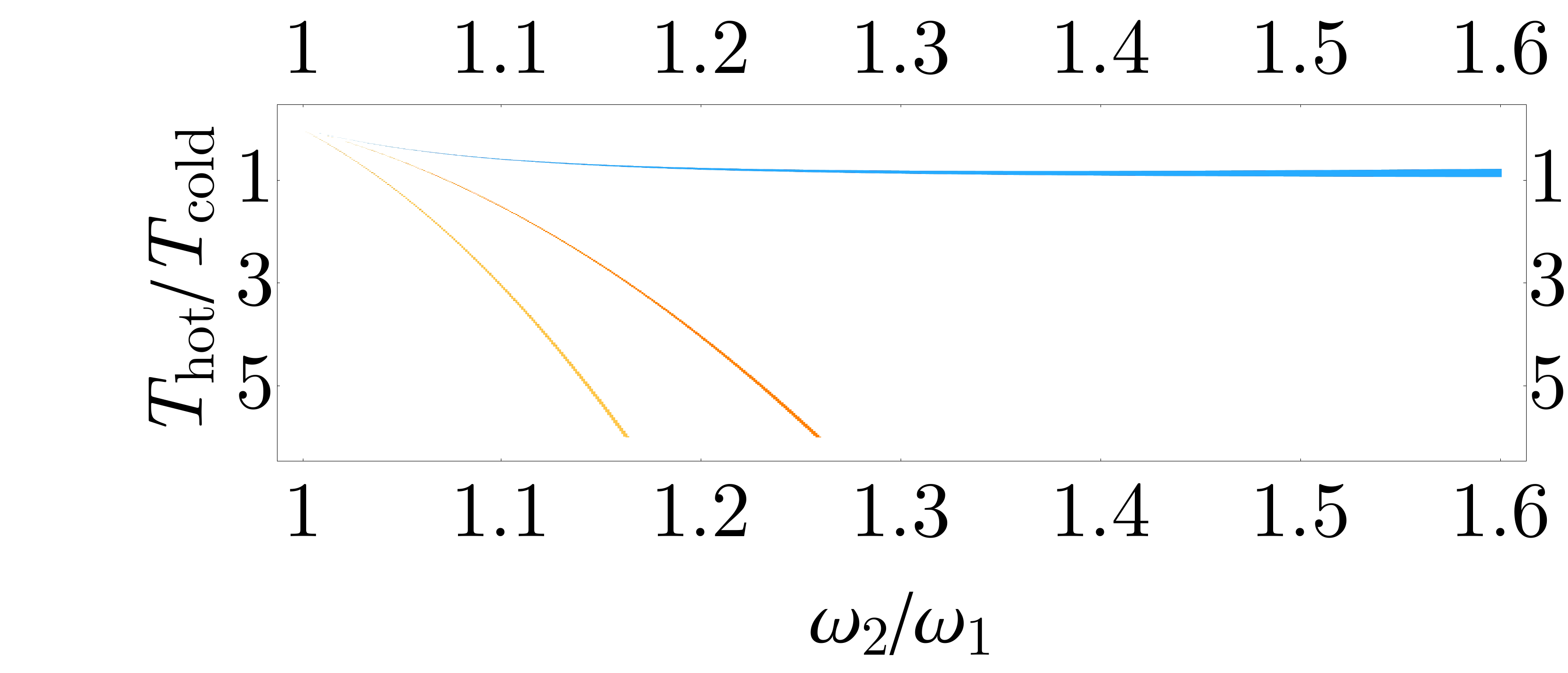}
}
\caption{\label{fig:op} Operational diagrams showing the differences in parameter space under which the cycles function as an engine (bottom, yellow region), heater I (middle, orange region), and refrigerator (top, blue region). (a) gives the difference of the boson and fermion parameter space, (b) gives difference between the boson and independent single particles, and (c) the difference between the the independent single particles and fermions. Parameters are $\hbar = k_B = 1$.}
\end{figure*}

\begin{figure*}
\centering
\subfigure[]{
\includegraphics[width=.3\textwidth]{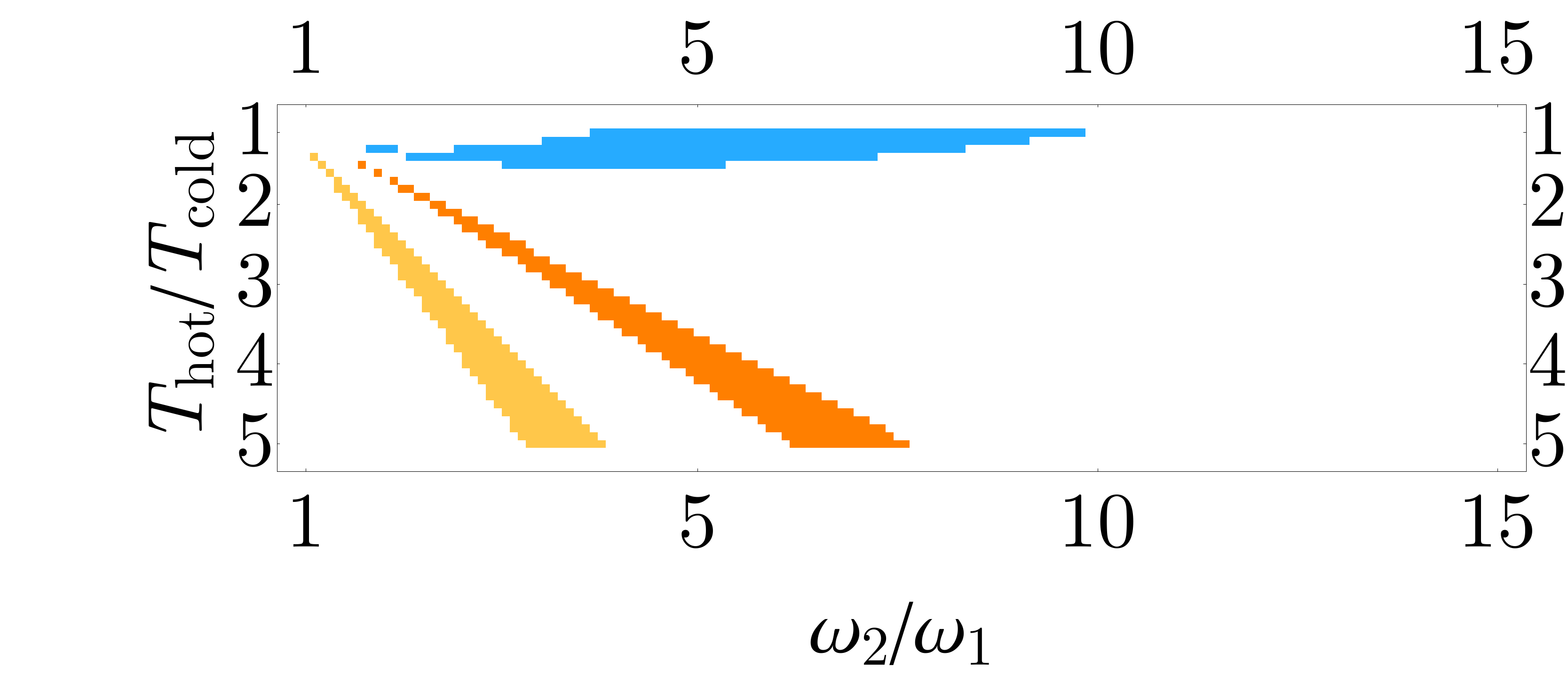}
}
\subfigure[]{
\includegraphics[width=.3\textwidth]{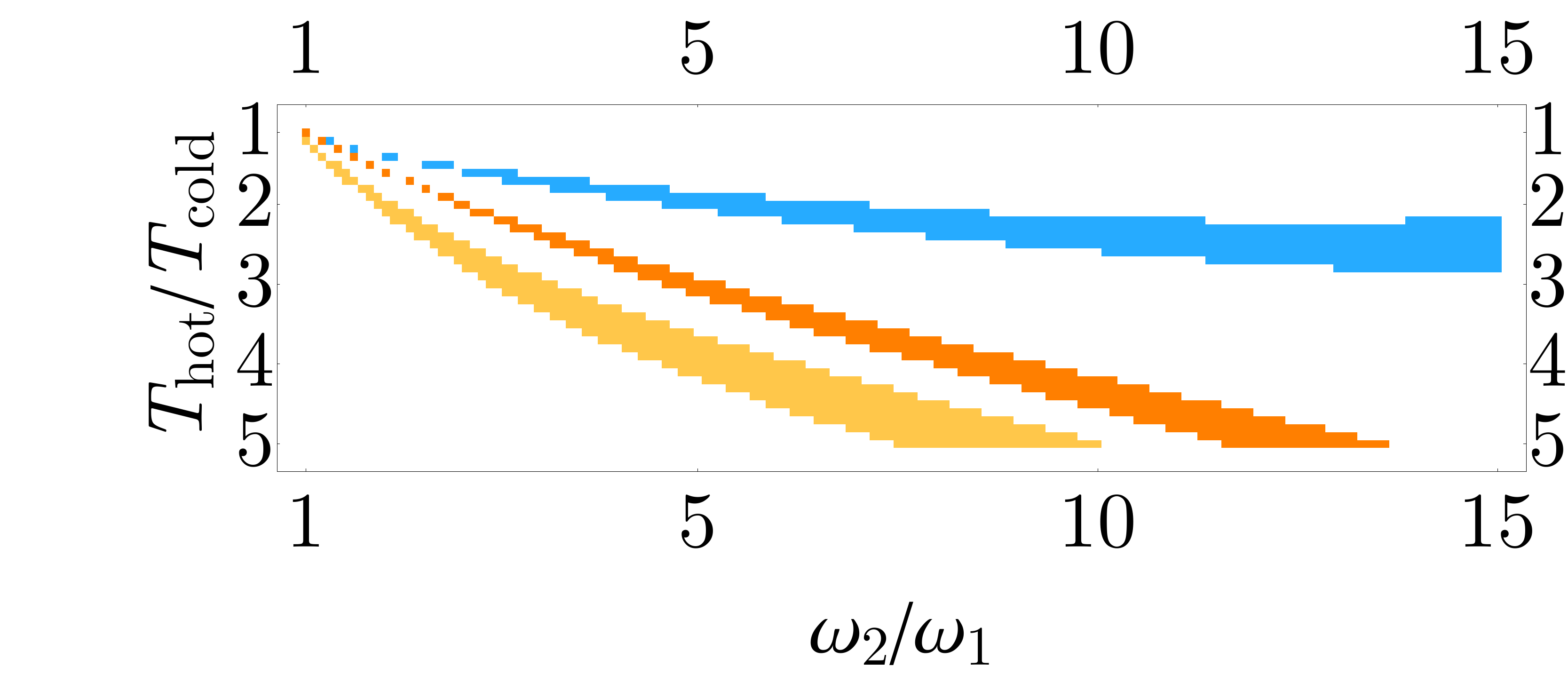}
}
\subfigure[]{
\includegraphics[width=.3\textwidth]{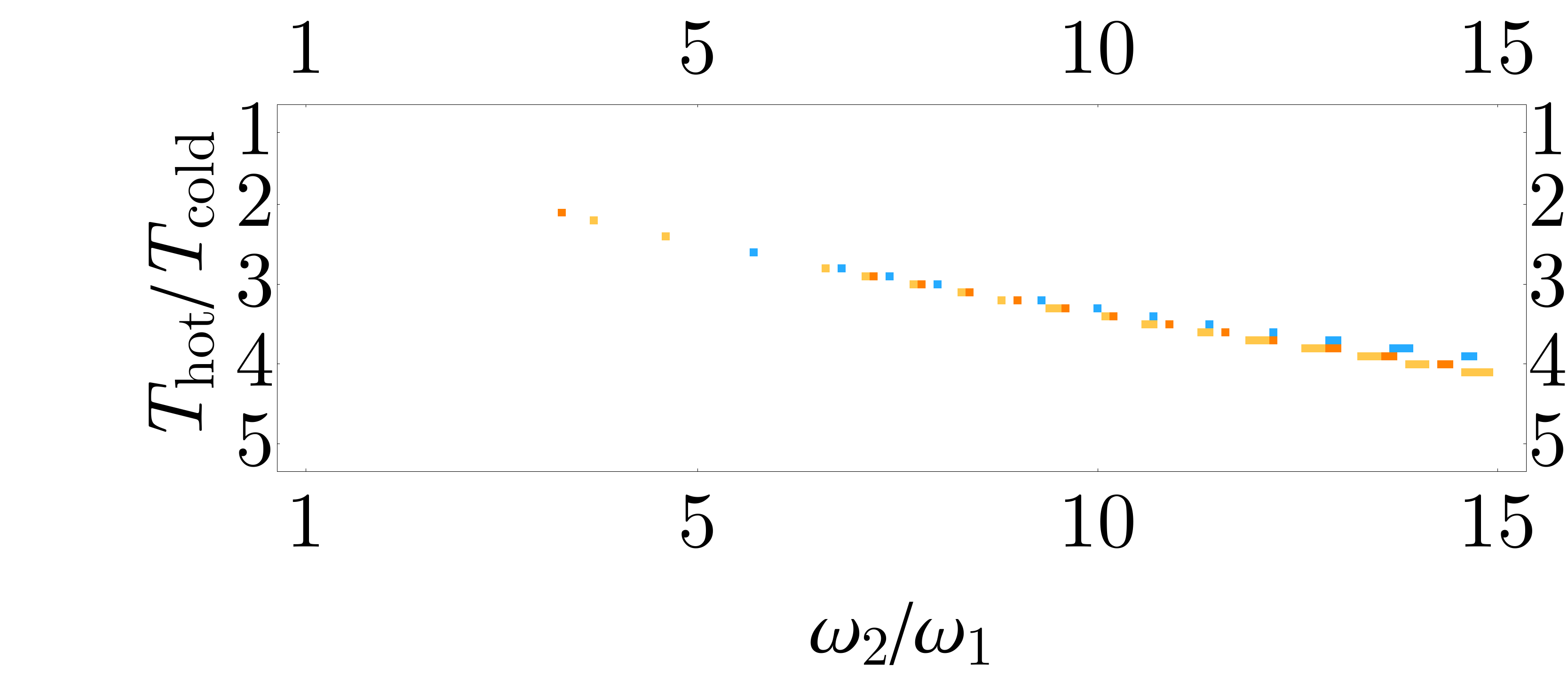}
}
\caption{\label{fig:lin} Operational diagrams showing the difference between the bosonic and fermionic engine regimes (bottom, yellow region), heater I regimes (middle, orange region) and refrigerator regimes (top, blue region) with the linear protocol for cycle times (a) $\tau = 1$, (b) $\tau = 10$, and (c) $\tau = 200$. Parameters are $\hbar = k_B = 1$.}
\end{figure*}

Aside from engine characterizations, the other main area in which a quantum advantage could manifest is in the size of the parameter space in which the cycle functions as the desired type of thermal machine. In general, there are four possible types of thermal machines allowed by the second law: engine, refrigerator, and two types of heater \cite{rana}. An engine extracts work from heat flowing between a hot and cold reservoir. A refrigerator puts in work in order to facilitate the flow of heat from a cold to a hot reservoir. The first type of heater puts in work to facilitate heat flow to both reservoirs. The second type of heater puts in work to facilitate the flow of heat from the hot to the cold reservoir. The operation of these machines is summarized in Fig.~\ref{fig:block}.

When thinking practically, heater II is clearly the least desired mode of operation as it involves spending work to facilitate a process that will occur spontaneously. With this in mind we define an advantage as an expansion of the parameter space under which the cycle functions as either engine, refrigerator, or heater I and a corresponding reduction in the parameter space where it functions as heater II.

By determining the signs for the work and heat components we can determine which thermal machine the cycle is functioning as for any combination of frequencies and bath temperatures. Figure~\ref{fig:op} gives a comparison between the operational space of the bosonic and fermionic systems as a function of their parameters. We see that the bosonic system experiences an expanded operational space for both the engine, refrigerator and heater I machines in comparison to both fermions (Fig.~\ref{fig:op}a) and an equivalent number of independent single particle engines (Fig.~\ref{fig:op}b). Conversely the fermionic system experiences a reduced operational space under the same comparisons (Fig.~\ref{fig:op}a and Fig.~\ref{fig:op}c). This demonstrates a clear advantage for the bosonic engine by our above definition.        

\subsection{Linear Protocol and Instantaneous Power}

It is also of interest to explore how the operational spaces evolve as the cycle transitions from the instantaneous sudden switch protocol to the infinite-time quasistatic limit. In order to do so we consider a linear protocol that allows us to interpolate smoothly between these limits,
\begin{equation}
\omega(t) = \left(\omega_1^2 + \delta\omega \frac{t}{\tau}\right)^{1/2}.
\end{equation}
A comparison between the the operational space of the bosonic and fermionic systems at selected cycle times, $\tau$, is shown in Fig.~\ref{fig:lin}. As cycle time increases the heater I and heater II regimes shrink while the engine and refrigerator regimes expand. We see for very large $\tau$, i.e. approximately quasi-static, the heater regimes disappear entirely and the boson and fermion operational diagrams become nearly identical. This matches expectations as the heaters are fundamentally non-equilibrium machines. The convergence of the operational diagrams is due to the fact that in the quasi-static limit both cycles performance approach the same limits (such as the Carnot and Curzon-Ahlborn efficiencies).             

In conclusion, we have seen in all explored characteristics that the bosonic working medium outperforms the fermionic one. The linear protocol allows us to examine how the internal energy and power evolve throughout the isentropic strokes which can provide us with insight into the source of this advantage.  

\begin{figure}
		\includegraphics[width=.48\textwidth]{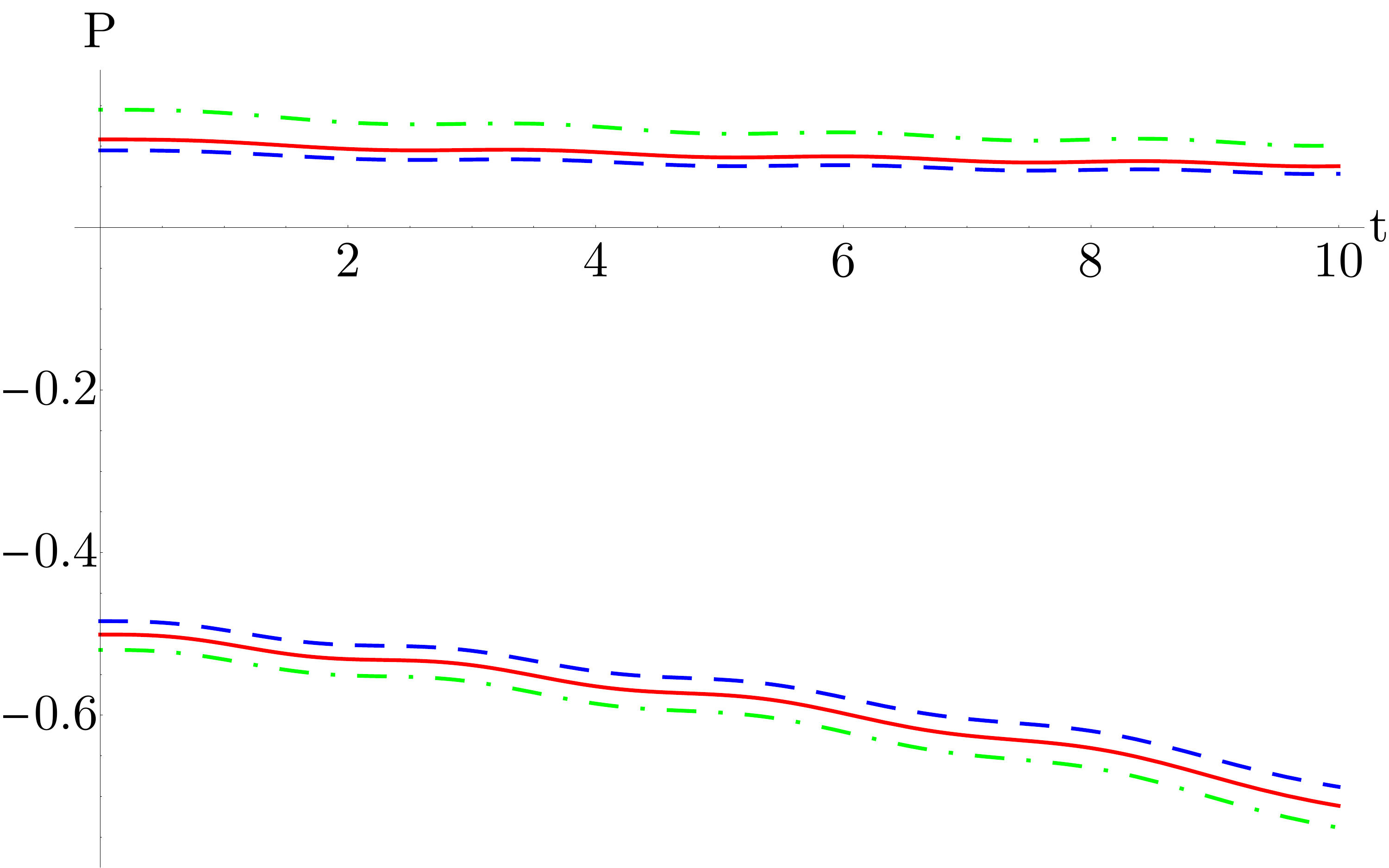}
		\caption{\label{fig:lin_power}Instantaneous power output during the compression (positive-valued lines) and expansion (negatively-valued lines) strokes for bosons (dashed, blue line), fermions (dot-dashed, green line), and the sum of two single particle engines (solid, red line). Note that in this case negative power represents work \textit{extraction} from the engine. We have taken $\hbar = k_B = 1$. Other parameters are $\omega_1 = 1$, $\omega_2=\sqrt{2}$, $\tau = 10$, $T_{\mathrm{cold}}=1$, and $T_{\mathrm{hot}}=10$.}
	\end{figure}

Figure~\ref{fig:lin_power} depicts the instantaneous power (the time derivative of the instantaneous internal energy) during both the expansion (opening) and compression (closing) strokes. We see that at any given time during the expansion stroke the bosons are \textit{extracting} less work than an equivalent number of independent single particle engines, but at the same time during the compression stroke they require less work \textit{input} resulting in a net gain in performance. The reverse is true for fermions. However, in their case the extra work input required outweighs the advantage they gain on extraction resulting in a net decrease in performance. Physically, this means that the repulsive force between fermions hinders more on compression than it aids in expansion, while the opposite is true of the attractive force between bosons. We also note that the slope of the instantaneous power and the separation between each case remains roughly constant (aside from some small oscillations induced by the form of $Q^*$ in the linear protocol). This is indication that the variation in performance arise solely from the differences in internal energy and not from differences in entropy induced by changes quantum correlations.

\section{Concluding Remarks}

In this work we have examined a quantum Otto engine operating on a two-particle working medium consisting of either spinless bosons or fermions through fully analytical models of the state dynamics. We have shown that in all examined engine characterizations, including efficiency, power, EMP, trade-off between efficiency and power, and operational parameter space, the bosonic system displays enhanced performance while the fermionic system displays reduced performance. This enhancement (or reduction) persists in comparison to the performance of an equivalent number of single-particle quantum engines clearly indicating that this effect arises from the particle symmetry. We have examined the time-dependent behavior of the instantaneous power output throughout the isentropic strokes for the case of a linear protocol and found that the origin of this effect lies in the differences in internal energy between the bosons and fermions that result from the Pauli exclusion principle.   

Wave-function symmetry is an inherently quantum property, as such this increase in performance (for bosons) is a demonstration of a truly quantum advantage. Beyond this however, the wave-function symmetry is also an additional \textit{information-bearing degree of freedom} \cite{Deffner2013} available to the system. Using information as a resource in a thermodynamic system, is an area that has seen much recent activity \cite{bengtwo, bengtsson, toyabe, park, hewgill, error, micadei, abahtwo, Levy, toytwo, Funo, Kim, Quan}. We leave an exploration of how this additional information resource may be leveraged as a topic to be explored in future work.          

\begin{acknowledgments}

It is a pleasure to thank Priyo Shankar Pal for informative discussion, especially in regards to the operational characterization of non-equilibrium thermal machines, as well as Akram Touil for enlightening conversation in regards to information measures in thermodynamic systems.  S.D. acknowledges support from the U.S. National Science Foundation under Grant No. CHE-1648973. This research was supported by grant number FQXi-RFP-1808 from the Foundational Questions Institute and Fetzer Franklin Fund, a donor advised fund of Silicon Valley Community Foundation (S.D). 

\end{acknowledgments}

\appendix

\onecolumngrid

\section{Density Operator Expressions}
\label{Appendix A}

This appendix provides the full expression for the two-particle thermal state \eqref{eq:thermal},     
\begin{equation}
\begin{split}
			\rho_0 &= \frac{1}{Z} \frac{m \omega}{2 \pi \hbar \sinh{(\beta \hbar \omega)}}
			\bigg[ e^{- \frac{m \omega}{4 \hbar}\{[(x_1+y_1)^2+(x_2+y_2)^2]\tanh{(\beta \hbar \omega)}+[(x_1-y_1)^2+(x_2-y_2)^2]\coth{(\beta \hbar \omega)}\}} \\
			&\pm e^{- \frac{m \omega}{4 \hbar}\{[(x_2+y_1)^2+(x_1+y_2)^2]\tanh{(\beta \hbar \omega)}+[(x_2-y_1)^2+(x_1-y_2)^2]\coth{(\beta \hbar \omega)}\}}\bigg].
\end{split}
\end{equation}

\section{Two Particle Propagator}
\label{Appendix B}

In this appendix we outline the derivation of the propagator in space representation for a two particle harmonic system and apply it to the thermal state density operator. 

For a given two particle state represented by density operator $\rho$ the propagator is defined by,
\begin{equation}
\label{eq:b1}
\rho_t(x_1,x_2,y_1,y_2) = \int dx_1^0 \int dx_2^0 \int dy_1^0 \int dy_2^0 
\,U_2 (x_1,x_1^0,x_2,x_2^0) \rho_0 (x_1^0,x_2^0,y_1^0,y_2^0) U_2^\dagger (y_1,y_1^0,y_2,y_2^0).
\end{equation}
Noting that in energy representation \cite{gong},
\begin{equation}
\bra{n_1 n_2} U_2 \ket{n_{1}^{0} n_{2}^{0}} = \frac{1}{2} \left[  \bra{n_1} U_1 \ket{n_{1}^{0}} \bra{n_2} U_1 \ket{n_{2}^{0}} 
\pm \bra{n_1} U_1 \ket{n_{2}^{0}} \bra{n_2} U_1 \ket{n_{1}^{0}} \right]\,, 
\end{equation}
and changing the basis in the expression \eqref{eq:b1} into energy representation we have,
\begin{equation}
U_2= \frac{1}{2} \sum\limits_{n_1,n_2 = 0}^\infty \braket{x_1 x_2}{n_1 n_2}
\left[ \sum\limits_{n_1^0,n_2^0 = 0}^\infty \left( \bra{n_1} U_1 \ket{n_1^0} \bra{n_2} U_1 \ket{n_2^0} 
\pm \bra{n_1} U_1 \ket{n_2^0} \bra{n_2} U_1 \ket{n_1^0} \right) \braket{n_1^0 n_2^0}{x_1^0 x_2^0} \right]\,,
\end{equation}
%and,
%\begin{align}
%\begin{split}
%U_2^\dagger (x_1,x_1^0,x_2,x_2^0) &= \frac{1}{2} \sum\limits_{m_1,m_2 = 0}^\infty \Bigg[ \sum\limits_{m_1^0,m_2^0 = 0}^\infty \braket{y_1^0 y_2^0}{m_1^0 m_2^0} 
%\Big( \bra{m_1^0} U_1^\dagger \ket{m_1} \bra{m_2^0} U_1^\dagger \ket{m_2} \\
%&\pm \bra{m_2^0} U_1^\dagger \ket{m_1} \bra{m_1^0} U_1^\dagger \ket{m_2} \Big) \Bigg] \braket{m_1 m_2}{y_1 y_2}. 
%\end{split}
%\end{align}
where,
\begin{equation}
\braket{x_1 x_2}{n_1 n_2} = \frac{1}{2} \left[ \psi_{n_1} (x_1) \psi_{n_2} (x_2) \pm \psi_{n_1} (x_2) \psi_{n_2} (x_1) \right].
\end{equation}

Further in position representation the harmonic oscillator energy eigenstates are,  
\begin{equation}
\psi_n(x)=\frac{1}{\sqrt{2^n n!}} \bigg(\frac{m \omega}{\pi \hbar} \bigg)^{1/4} e^{- \frac{m \omega x^2}{2 \hbar}} H_n \bigg( \sqrt{\frac{m \omega}{\hbar}}x\bigg)\,,
 \end{equation}
and the orthogonality condition of the Hermite polynomials is,
\begin{equation}
\sum\limits_{n=0}^\infty \frac{1}{\sqrt{2^n n!}} H_n (x) H_n (y) = \sqrt{\pi} e^{\frac{1}{2} (x^2 + y^2)} \delta(x-y),
\end{equation}
where $\delta(x-y)$ is the Dirac delta. Thus, we obtain
\begin{equation}
U_2(x_1,x_1^0,x_2,x_2^0) = \frac{1}{2} \left[  U_1(x_1,x_1^0) U_1(x_2,x_2^0)
\pm U_1(x_1,x_2^0) U_1(x_2,x_1^0)\right]. 
\end{equation} 
with the same method allowing for the simplification of the conjugate to a similar expression.

Applying the two particle propagator to the thermal state \eqref{eq:b1} yields the full time-evolved density operator,     
\begin{equation}
\begin{split}
&\rho_t(x_1,x_2,y_1,y_2)= \frac{m \omega}{2 \pi \hbar (Y_t^2+X_t^2 \omega^2)} \left(e^{\mp \beta \hbar \omega}-1\right) \\
&\quad\times\bigg\{ e^{\frac{m}{2 \hbar (Y_t^2+X_t^2 \omega^2)}\left[i (x_1^2+x_2^2-y_1^2-y_2^2)(Y_t \dot{Y}_t+X_t\dot{X}_t\omega^2)-\omega (x_1^2+x_2^2+y_1^2+y_2^2) \mathrm{coth}(\beta \hbar \omega)
+2 \omega (x_1 y_1 +x_2 y_2) \mathrm{csch}(\beta \hbar \omega) \right]} \\
&\quad\pm e^{\frac{m}{2 \hbar (Y_t^2+X_t^2 \omega^2)}\left[i (x_1^2+x_2^2-y_1^2-y_2^2)(Y_t \dot{Y}_t+X_t\dot{X}_t\omega^2)
- \omega(x_1^2+x_2^2+y_1^2+y_2^2) \mathrm{coth}(\beta \hbar \omega)+2 \omega (x_2 y_1 +x_1 y_2) \mathrm{csch}(\beta \hbar \omega) \right]} \bigg\}.
\end{split}
\end{equation}
Here the top sign denotes the boson state and the the bottom sign denotes the fermion state.

\section{Wigner Formalism}
\label{Appendix C}

Finally, we outline the derivation of the thermal and time-evolved state Wigner distributions. The definition of the Wigner distribution generalized to a two particle density matrix is,
\begin{equation}
W(x_1,p_1,x_2,p_2) = \frac{1}{4 \pi^2 \hbar^2} \int du_1 \int du_2\,
\rho\left(x_1+\frac{u_1}{2},x_2+\frac{u_2}{2},x_1-\frac{u_1}{2},x_2-\frac{u_2}{2}\right)
e^{-\frac{i p_1 u_1}{\hbar}}e^{-\frac{i p_2 u_2}{\hbar}}.
\end{equation}
Plugging in the thermal state we obtain the thermal state Wigner function,
\begin{equation}
\begin{split}
&W_0(x_1,p_1,x_2,p_2) = \frac{\mathrm{sech}^2\left(\beta \hbar \omega/2 \right)}{\pi^2 \hbar^2 (\mathrm{csch}^2\left( \beta \hbar \omega/2 \right) \pm 2 \mathrm{csch}\left(\beta \hbar \omega \right))}\\
&\quad\times\left( e^{-\frac{(p_1^2 +p_2^2 +m^2(x_1^2+x_2^2)\omega^2)\mathrm{tanh}\left(\beta \hbar \omega/2 \right)}{m \omega \hbar}}\pm 2 e^{\frac{-(p_1^2+p_2^2 +m^2(x_1^2+x_2^2)\omega^2)\mathrm{coth}(\beta \hbar \omega)+2(p_1 p_2 +m^2 \omega^2 x_1 x_2)\mathrm{csch}(\beta \hbar \omega)}{m \omega \hbar} }\right)\,.
\end{split}
\end{equation}
Here the top sign denotes the boson distribution, and the bottom sign the fermion. Integrating over the momentum coordinates yields the position-space probability distribution, plotted in Fig.~\ref{fig:thermal}, 
\begin{equation}
\begin{split}
&P_x(x_1,x_2) = \frac{2 m \omega }{\pi \hbar }\frac{\mathrm{csch}\left(\beta \hbar \omega\right)}{\mathrm{csch}^2\left(\beta \hbar \omega/2 \right) \pm 2 \mathrm{csch}\left(\beta \hbar \omega \right)}\\
&\quad\times \left(e^{-\frac{m \omega}{\hbar} (x_1^2 + x_2^2) \mathrm{tanh}\left(\beta \hbar \omega/2\right)} \pm e^{-\frac{m \omega}{\hbar} (-2 x_1 x_2+(x_1^2 +x_2^2)\mathrm{cosh}\left(\beta \hbar \omega\right))\mathrm{csch}\left(\beta \hbar \omega\right)}\right).
\end{split}
\end{equation}

Repeating the same process for the time evolved density operator, given in (9), yields the time-evolved Wigner distribution,
\begin{equation}
\label{eq:W_evolved}
\begin{split}
&W_t(x_1,p_1,x_2,p_2) = \frac{1}{2 \pi^2 \hbar^2} \bigg[e^{-\frac{[p_1^2 \alpha^2 +p_2^2 \alpha^2 - 2 m \alpha \gamma (p_1 x_1 + p_2 x_2) +m^2(x_1^2 +x_2^2)(\omega^2 + \gamma^2)]\mathrm{tanh}[\frac{1}{2}\beta \hbar \omega]}{\alpha m \omega \hbar}}(\pm 1 \mp e^{\mp \beta \hbar \omega})\mathrm{tanh}\left(\beta \hbar \omega/2\right) \\
&\quad\mp e^{-\frac{\mathrm{coth}(\beta \hbar \omega)(p_1^2 \alpha^2 +p_2^2 \alpha^2 -2 m \alpha \gamma(p_1 x_1 + p_2 x_2)+m^2(x_1^2+x_2^2)(\omega^2+\gamma^2)-2\{p_1 \alpha [Y_t (p_2 Y_t-m x_2 \dot{Y}_t)+X_t (p_2 X_t-m x_2 \dot{X}_t)\omega^2]+m x_1[-p_2 \alpha \gamma+m x_2 (\omega^2+\gamma^2)]\}\mathrm{sech}(\beta \hbar \omega))}{\alpha m \omega \hbar}}\\
&\quad(\mp 1\pm e^{\mp \beta \hbar \omega})\bigg], 
\end{split}
\end{equation}
where $\alpha = Y_t^2 + X_t^2 \omega^2$ and $\gamma = Y_t \dot{Y}_t + X_t \dot{X}_t \omega^2$. 

%In the Wigner function formalism the expectation value of an operator $A$ is given by,
%
%\begin{equation}
%\la A \ra = \int \mathrm{d}x_1 \int \mathrm{d}x_2 \int \mathrm{d}p_1 \int \mathrm{d}p_2 \;  W_t(x_1,p_1,x_2,p_2) \tilde{A}(x_1,p_1,x_2,p_2).
%\end{equation}   
%Here $\tilde{A}(x_1,p_1,x_2,p_2)$ is the Weyl transform of operator $A$. In the case that $A$ is a function of the position and momentum observables, $\tilde{A}$ is identical in structure to the $A$, however, instead of being dependent on non-commuting \textit{operators} $x$ and $p$, it is simply a function of $x$ and $p$ \textit{variables}, which can commute freely \cite{case}. Plugging in
%
%\begin{equation}
%\tilde{H}(x_1,p_1,x_2,p_2) = \frac{p_1^2}{2m} + \frac{p_2^2}{2m} +\frac{1}{2} m \omega_t^2 (x_1^2 + x_2^2)
%\end{equation}
%with the proper values of $\beta, \omega,$ and $\omega_t$ for the desired point in the cycle and carrying out the necessary integrals yields the internal energies given in Eq. (9).    

\twocolumngrid

\bibliography{refs}
		
\end{document}